\documentclass[manuscript]{acmart}
\AtBeginDocument{%
  }

\setcopyright{acmlicensed}
\copyrightyear{2018}
\acmYear{2018}
\acmDOI{XXXXXXX.XXXXXXX}
\acmConference[Conference acronym 'XX]{Make sure to enter the correct
  conference title from your rights confirmation email}{June 03--05,
  2018}{Woodstock, NY}
\acmISBN{978-1-4503-XXXX-X/2018/06}

\begin{document}

\title{Touvigation: Embodied Adaptive Object Acquisition for Blind and Low-Vision Users in Unfamiliar Indoor Environments}

\author{George Xi Wang}
\affiliation{%
  \institution{Stony Brook University}
  \city{Stony Brook}
  \state{New York}
  \country{United States}}
\email{george.x.wang@stonybrook.edu}
\affiliation{%
  \institution{New York University}
  \city{Brooklyn}
  \state{New York}
  \country{United States}}
\email{xw3617@nyu.edu}

\author{Xiangyu Li}
\affiliation{%
  \department{Department of Computer Science}
  \institution{Brown University}
  \city{Providence}
  \state{Rhode Island}
  \country{United States}}

\author{Shaoyue Wen}
\affiliation{%
  \institution{Imperial College London}
  \city{London}
  \country{United Kingdom}}
\email{jw7525@ic.ac.uk}

\author{Jiaqian Hu}
\affiliation{%
  \department{Translation and Localization Management}
  \institution{Middlebury Institute of International Studies at Monterey}
  \city{Monterey}
  \state{California}
  \country{United States}}
\email{jiaqianh@middlebury.edu}

\author{Junan Xie}
\affiliation{%
  \department{Internet of Things Thrust}
  \institution{The Hong Kong University of Science and Technology (Guangzhou)}
  \city{Guangzhou}
  \country{China}}
\email{jxie622@connect.hkust-gz.edu.cn}

\author{Yupeng Wang}
\affiliation{%
  \institution{Tongji University}
  \city{Shanghai}
  \country{China}}

\author{Ziyue Shi}
\affiliation{%
  \institution{Shanghai Qibao Dwight High School}
  \city{Shanghai}
  \country{China}}

\author{Qijun Chen}
\affiliation{%
  \department{College of Electronic and Information Engineering}
  \institution{Tongji University}
  \city{Shanghai}
  \country{China}}

\author{Maaike Bouwmeester}
\affiliation{%
  \institution{New York University}
  \city{New York City}
  \state{New York}
  \country{United States}}
\email{mb262@nyu.edu}

\author{Yuhua Jin}
\authornote{Co-corresponding authors.}
\affiliation{%
  \department{School of Science and Engineering}
  \institution{The Chinese University of Hong Kong, Shenzhen}
  \city{Shenzhen}
  \state{Guangdong}
  \country{China}}
\email{yuhuajin@cuhk.edu.cn}

\author{Jing Qian}
\authornotemark[1]
\affiliation{%
  \department{College of Electronic and Information Engineering}
  \institution{Tongji University}
  \city{Shanghai}
  \country{China}}

\authorsaddresses{}
\renewcommand{\shortauthors}{Wang et al.}

\begin{abstract}

Blind and low-vision users benefit from AI object-search systems to assist movement and object-searching in unfamiliar spaces. However, existing systems have high latency and require BLV users to mentally translate AI instructions. Through formative interviews with eight BLV participants, we identified existing challenges including AI guidance misaligned with actions, high latency, and lacking tactile identification. We present Touvigation, a phone-based object-acquisition system combines an LLM with local 3D reconstruction for low-latency embodied guidance. Touvigation uses a multi-stage reference switching method for end-to-end object acquisition. An empirical study with 12 BLV participants comparing Touvigation, Doubao (an MLLM AI Agent) for success rate, completion time, and workload in 2 different settings found our system reached 100\% success rate (58\% for Doubao and 85\% for unassisted), significantly reduced the overall cognitive load and mental demand with the fastest completion experience. Subjective ratings reflect high scores for Touvigation's spatial awareness, trust, and perceived safety.

\end{abstract}





\begin{teaserfigure}
  \centering
  \includegraphics[width=1\textwidth]{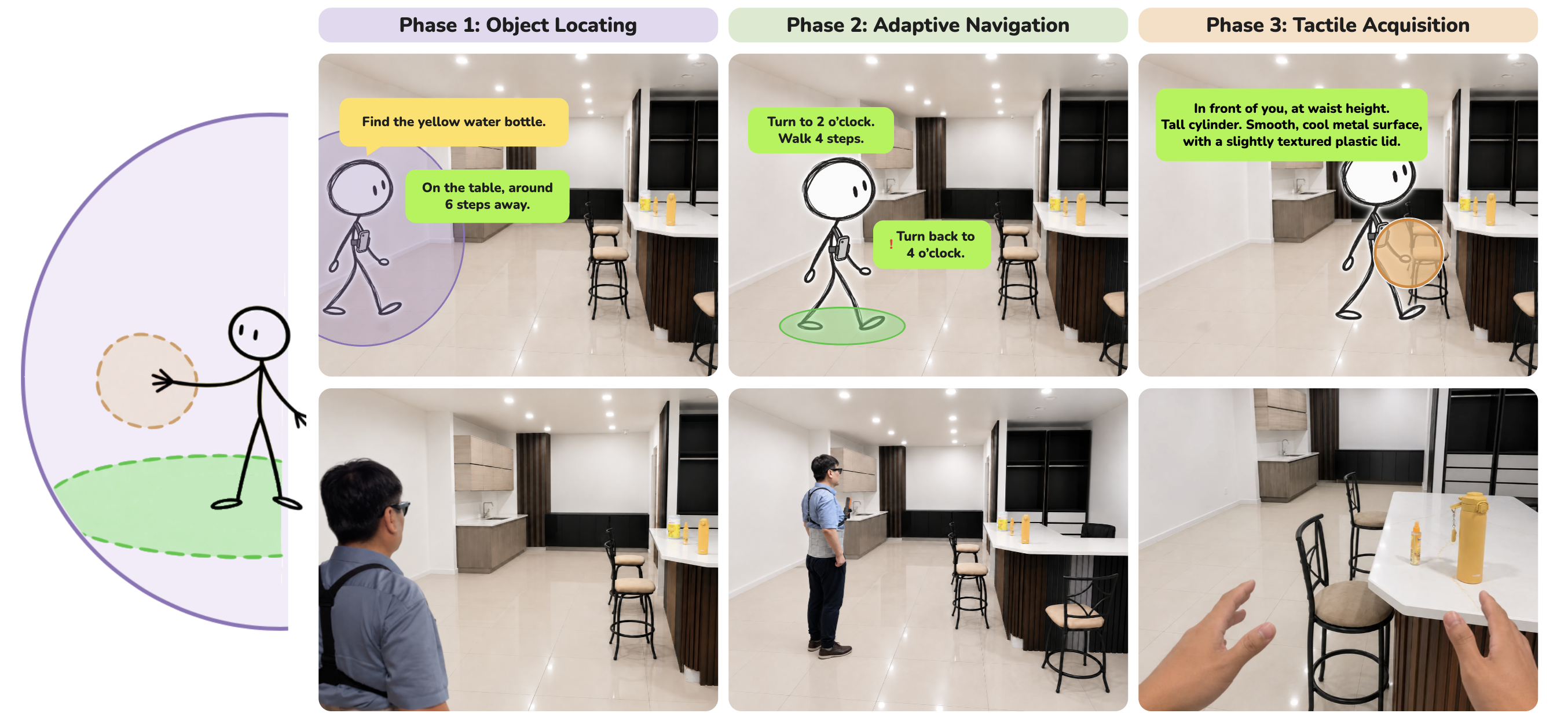}
  \caption{Touvigation supports hands-free object acquisition through a chest-mounted, LiDAR-equipped iPhone. The illustrated interaction progresses through object locating, adaptive navigation, and tactile verification. Spoken guidance uses body-relative clock directions, personalized step counts, and descriptions of target height and tactile features to help users locate, reach, and identify the requested object.}
  \label{fig:design-space}
  \Description{The figure illustrates Touvigation’s three-stage process for helping a blind or low-vision user locate and physically acquire an object. On the left, a simplified diagram shows a person surrounded by three body-centered guidance regions: a large lavender area representing orientation to the surrounding space, a green region around the feet representing walking guidance, and an orange region around the extended hand representing reaching guidance. On the right, three columns show the process in the same indoor room, with a yellow water bottle on a counter as the target. Phase 1, Object Locating: the system identifies the target and gives broad spatial guidance. A chest-mounted phone user receives the instructions, “Find the yellow water bottle” and “On the table, around 6 steps away.” Phase 2, Adaptive Navigation: guidance becomes movement-specific as the user approaches the target. Instructions include “Turn to 2 o’clock. Walk 4 steps,” followed by a corrective instruction, “Turn back to 4 o’clock,” illustrating that guidance is updated when the user changes direction. Phase 3, Tactile Acquisition: once the user is within reach, the system shifts from walking guidance to hand-level guidance. It says, “In front of you, at waist height. Tall cylinder. Smooth, cool metal surface, with a slightly textured plastic lid.” The final first-person image shows both hands reaching toward the yellow bottle on the counter.}
\end{teaserfigure}

\received{20 February 2007}
\received[revised]{12 March 2009}
\received[accepted]{5 June 2009}

\maketitle

\section{Introduction}
Visual assistance provides a new way for blind and low-vision (BLV) users to move around a space~\cite{bigham2010vizwiz, microsoft2017seeingai, google2025lookout, orcam2024myeye3pro, apple2023accessibilityfeatures} as modern VLM-based natural-language systems speaks great amount of details about surrounding descriptions~\cite{bemyeyes2023bemyai, gonzalez2026multimodal, google2024talkbackgemininano}, helping BLV users to find objects in everyday settings. These systems are especially important and useful when a BLV user enters a foreign, or public space. However, acquiring an object requires more than identifying it; users must locate, approach, and interact with the target~\cite{liu2026objectfinder,sridhar2025navisense}. A description such as ``the cup is on the table'' leaves users to determine how to approach it~\cite{zamiri2026say, liu2026objectfinder, chang2025probing}, position their hand~\cite{rahman2023takemyhand}, and distinguish it from nearby objects~\cite{chang2026touchscribe, kacorri2017people}. As users move, guidance needs to update continuously to remind users of spatial information such as accessible route or obstacles, and stays connected with their bodies, or in an embodied manner ~\cite{dourish2001action,klemmer2006bodies, qian2025duozone}.

Providing such guidance requires a system to understand the spatial semantics, respond in a timely with accuracy~\cite{chang2025probing,kuribayashi2026time}, while being able to guide the user to the object from beginning to end (consistency). Yet prior work suffer from hallucinations~\cite{xie2025beyond,adnin2024look,alharbi2024misfitting, azaria2023internal}, difficulty to remain consistent on 3D spatial distances~\cite{chen2024spatialvlm,ma20253dsrbench}. Targets may leave the camera's field of view during approach~\cite{manduchi2014last,liu2026objectfinder}, undermining the tracking consistency. Most importantly, current systems that rely on semantic descriptions suse MLLMs that suffer from network delays, delivering out-of-sync guidance for users~\cite{romero2026navigation,lin2026multimodal}. Beyond tracking, most existing systems help users navigate with simple text or voice descriptions, which can confuse BLV users as interaction distances change as they approach the object. How can we implement a low-latency, consistent, and easy-to-follow system that helps BLV users to acquire objects in a foreign space?

We present \textbf{Touvigation}, a system that uses a multi-stage embodied description design to offer non-visually reliant descriptions by integrating Simultaneous Localization and Mapping (SLAM) with VLMs for swift, accurate contextual guidance. Based on the system's capability to reconstruct the 3D object's location in realtime tracking, it uses a two-stage guides that first bring the user to the target object within their arm's reach, and then enabling detailed semantic environment descriptions with a ``touch'' route to help acquiring the object (Figure~\ref{fig:design-space}). A 12 blind participant within-subject study was performed in two unfamiliar rooms showed that our system succeeded in all target acquiring trials, when the controlled condition Doubao and unassisted search scored 58\% and 85\% respectively. Our system further significantly reduced the time to acquire the objects and overall cognitive load compared to both other conditions. Semi-structured interviews revealed that multi-stage embodied cues helped participants translate spatial information into executable movements while tactile descriptions supported target identification. Participants also reported greater overall perceived safety and trust when using Touvigation.

We hope this work inspires a shift from merely describing the visual world to making it directly actionable for BLV users.Our \textbf{main contributions} are:
\begin{enumerate}
    \item a multi-stage embodied guidance paradigm that translates spatial information into actionable instructions spanning navigation and object acquiring,
    \item a SLAM-VLM architecture that uses a voting mechanism to combine semantic interpretation with persistent local 3D spatial tracking for continuously updated guidance,
    \item formative results of into BLV users' action needs and a twelve-BLV-user empirical evaluation demonstrating improved object acquisition success, completion time, cognitive load and perceived safety.
\end{enumerate}

\section{Related Work}
\label{sec:related}

\subsection{AI-Mediated Visual Access and Object Search}
Visual assistance for blind and low-vision (BLV) people has often been framed as a problem of making visual information available through answers, labels, and descriptions. VizWiz showed the value of near-real-time visual question answering for blind users~\cite{bigham2010vizwiz}, and later work characterized the everyday visual challenges and visual-question datasets that shaped this research area~\cite{brady2013visual,gurari2018vizwiz,zeng2020vision}. Recent MLLM-enabled tools extend this direction from single visual questions toward conversational scene interpretation, but studies with BLV users show recurring problems: users need more control over what is described, must verify errors and hallucinations, and cannot always turn fluent descriptions into action~\cite{gonzalez2024investigating,gonzalez2025towards,gonzalez2026multimodal,zamiri2026say,tang2025my,alharbi2024misfitting, mathis2025lifeinsight}. Work probing live video AI assistants further shows that these systems can help with static scenes but struggle with dynamic real-world assistance, timing, spatial precision, and risky assumptions about users' visual abilities~\cite{chang2025probing, chang2024worldscribe}.

Object-search systems are the closest part of this literature because they connect recognition with spatial localization. Kacorri et al. studied the feasibility and challenges of BLV users training personal object recognizers~\cite{kacorri2017people}, and Find My Things built this idea into a teachable AI system for finding personal objects~\cite{morrison2023understanding, wen2024find}. ObjectFinder supports open-vocabulary object search with interactive descriptions and navigation cues~\cite{liu2026objectfinder}. NaviSense combines conversational AI, LiDAR, and real-time audio-haptic feedback for object retrieval~\cite{sridhar2025navisense}. TouchScribe also targets hand-object interaction, using live visual descriptions to enrich what BLV users can learn while touching objects~\cite{chang2026touchscribe}. Touvigation differs from this line by treating object acquisition as a continuous embodied control problem: after semantic recognition ~\cite{zhang2025enhancing}, it keeps the target grounded in local 3D space and updates guidance as the user moves from room-scale locomotion to hand-scale reaching. Table~\ref{tab:system-comparison} compares Touvigation with existing systems.

\begin{table}[t]
  \caption{Comparison of Touvigation with existing visual-assistance, navigation, and object-search systems for BLV users. \checkmark{} = supported; -- = not supported.}
  \label{tab:system-comparison}
  \footnotesize
  \setlength{\tabcolsep}{4pt}
  \begin{tabular}{@{}l*{7}{c}@{}}
    \toprule
    System
      & Hands-free
      & \begin{tabular}[c]{@{}c@{}}Obstacle\\avoidance\end{tabular}
      & \begin{tabular}[c]{@{}c@{}}Target\\persists\\outside FOV\end{tabular}
      & \begin{tabular}[c]{@{}c@{}}Hand /\\reach\\guidance\end{tabular}
      & \begin{tabular}[c]{@{}c@{}}Tactile /\\material\\information\end{tabular}
      & \begin{tabular}[c]{@{}c@{}}Low-latency\\real-time\end{tabular}
      & \begin{tabular}[c]{@{}c@{}}Adapts to\\user's sensing\\strategy\end{tabular} \\
    \midrule
    Be My Eyes / Be My AI~\cite{bemyeyes2026app,bemyeyes2023bemyai} & --         & \checkmark & --         & \checkmark & --         & --         & --         \\
    Google Lookout~\cite{google2025lookout}                          & --         & \checkmark & --         & --         & --         & \checkmark & --         \\
    Doubao~\cite{bytedance2026doubao}                                & --         & \checkmark & --         & --         & --         & --         & --         \\
    Shike~\cite{shike2026navigation}                                 & \checkmark & --         & --        & --         & --         & \checkmark & --         \\
    Take My Hand~\cite{rahman2023takemyhand}                         & \checkmark & --         & --         & \checkmark & --         & \--  & --         \\
    NaviSense~\cite{sridhar2025navisense}                            & --         & --         & \checkmark & \checkmark & --         & -- & --         \\
    NaviGPT~\cite{zhang2025enhancing}                                & --         & \checkmark & --        & --         & --         & -- & --         \\
    ObjectFinder~\cite{liu2026objectfinder}                          & \checkmark & \checkmark & --         & --         & --         & -- & --         \\
    Seeing with the Hands~\cite{teng2025seeing}                      & \checkmark & --         & --         & \checkmark & \checkmark & -- & --         \\
    \midrule
    \textbf{Touvigation (ours)}                                      & \checkmark & \checkmark & \checkmark & \checkmark & \checkmark & \checkmark & \checkmark \\
    \bottomrule
  \end{tabular}
\end{table}

\subsection{Navigation and Wayfinding for BLV Travelers}
Navigation systems for BLV travelers have studied how to support route following, orientation, obstacle awareness, and exploration. The Last Meter examined visual guidance to a nearby target~\cite{manduchi2014last}, while NavCog and NavCog3 demonstrated large-scale indoor navigation with localization infrastructure and semantic environmental information~\cite{ahmetovic2016navcog,sato2019navcog3}. Studies of turn-by-turn navigation show that rotation instructions are themselves error-prone without vision~\cite{ahmetovic2018turn}. Other systems use wearable, robotic, or environmental sensing: Headlock helps cane users cross large open spaces~\cite{fiannaca2014headlock}, CaBot explores autonomous robot guidance~\cite{guerreiro2019cabot, wang2026feelium}, RouteNav supports blind travelers in a transit hub~\cite{ren2023experiments}, StreetNav repurposes street cameras for precise outdoor navigation~\cite{jain2024streetnav}, and WanderGuide supports map-less exploration with a robotic guide~\cite{kuribayashi2025wanderguide}. Recent work also emphasizes that accessible navigation must be resilient to disruption rather than assuming clean routes, stable infrastructure, or uninterrupted sensing~\cite{cross2026resilience,parker2021wayfinding}.

This literature makes clear that BLV navigation is not only a path-planning problem. Guidance must be timely, localizable, and compatible with users' mobility practices. Most prior systems target movement through buildings, streets, museums, or transit spaces; Touvigation targets the smaller but demanding transition from moving through an unfamiliar room to acquiring a specific object. This transition changes the relevant action scale, so the guidance must shift from direction and steps to reachable surfaces, hand position, height, and tactile confirmation.

\subsection{Embodied and Nonvisual Spatial Guidance}
Embodied interaction argues that interaction is shaped by bodily action, risk, practice, and situated movement rather than abstract information alone~\cite{klemmer2006bodies, froese2009enactive}. Accessibility research has applied this principle through auditory, tactile, and multimodal guidance ~\cite{bharadwaj2019comparing,brewster2004tactons}. Dynamic audio can support peripersonal reaching~\cite{wilson2016peripersonal}, sonification can guide manual tasks~\cite{guarese2022microguidance}, and auditory hand-steering work studies how BLV users can follow eyes-free 3D hand paths~\cite{abe2025auditory,abe2026handsteering}. In virtual environments, auditory and haptic white-cane simulations show how nonvisual cues can make complex spatial layouts navigable~\cite{siu2020virtual, zhao2018enabling, tzovaras2009interactive, lecuyer2003homere}.

Several systems focus directly on hand-scale or object-scale guidance. AIGuide uses augmented reality to guide hand movement in a visual prosthetic context~\cite{lee2022aiguide, qin2026making}; Take My Hand explores automated hand-based spatial guidance using a miniature finger-mounted robot~\cite{rahman2023takemyhand}; LiSee uses headphone-based sensing to help users reach surrounding objects~\cite{chen2022lisee}; and TouchPilot guides blind users in learning complex 3D structures through touch~\cite{wang2023touchpilot}. These works show that the form of feedback matters as much as the spatial estimate. Touvigation builds on them by connecting hand-scale feedback to a preceding locomotion loop: the system first maintains a persistent target and computes a safe approach, then re-anchors instructions to the user's hand and nearby tactile landmarks when the object becomes reachable.

\section{Formative Study}
\label{sec:formative}
\begin{table}[t]
  \caption{Participant demographics from the formative interview study ($N=8$). Vision impairment levels follow the Chinese national visual-disability classification standard~\cite{gbt26341}, where Level~1 indicates more severe vision loss and Level~4 indicates lower severity.}
  \label{tab:formative-participants}
  \small
  \begin{tabular}{@{}llcclll@{}}
    \toprule
    Participant & Gender & Age & Vision Impairment Level & Onset of Impairment & Occupation  \\
    \midrule
    FP1 & Male   & 30 & 1 & Since age 4 & Student           \\
    FP2 & Female & 30 & 1 & Since 2021  & N/A              \\
    FP3 & Female & 27 & 4 & Since birth & Teacher           \\
    FP4 & Male   & 19 & 4 & Since age 9 & Student           \\
    FP5 & Female & 25 & 1 & Since 2023  & N/A               \\
    FP6 & Male   & 23 & 1 & Since age 6 & Massage Therapist  \\
    FP7 & Male   & 31 & 2 & Since birth & Massage Therapist \\
    FP8 & Male   & 24 & 3 & Since age 3 & Student            \\
    \bottomrule
  \end{tabular}
\end{table}

We conducted a formative interview study to examine how BLV participants used current AI visual-assistance tools in unfamiliar indoor spaces. 

\subsection{Participants}
We recruited 8 BLV participants (FP1--FP8) through a social media post on RedNote~\cite{xiaohongshu2026rednote}.Eligibility required participants to use AI-based visual assistance in their daily lives. Participants were 19--31 years old ($M=26.1$, $SD=4.2$) and varied in education, occupation, onset of impairment, and impairment level. Impairment ranged from Level~1 (most severe) to Level~4 under the Chinese national visual-disability classification standard~\cite{gbt26341}; Table~\ref{tab:formative-participants} provides individual details. All eight participants completed the study and were included in the analysis. The study was approved by our institutional review board.

\subsection{Data Collection and Analysis}
Each participant completed a remote semi-structured interview using Zoom or Tencent Meeting. Interviews lasted 30--60 minutes ($M=45.9$, $SD=10.7$) and were conducted in Mandarin. With participants' consent, the interviews were audio-recorded and transcribed into English for analysis. Three researchers coded the transcripts. They grouped the codes through affinity diagramming and refined the groupings through discussion until the team agreed on the five challenges reported below.

\subsection{Procedure}
The interview comprised three parts. It began with questions about participants' backgrounds and prior use of AI tools, including which tools they used, how they used them, and the tasks they completed with them. Participants then completed a think-aloud scenario imagining they had just entered a hotel room. The scenario involved two tasks: finding the remote to turn on the air conditioner and throw a piece of garbage to a trash can. For each task, participants described step by step how they would proceed \textit{without AI tools} and then \textit{with the tools} they currently used. The final part invited an open-ended reflection on what guidance, interactions, and environmental information they would want from a navigation system.

\subsection{Challenges in Current LLM-Powered Tools}
\begin{figure*}[t]
  \centering
  \includegraphics[width=0.8\linewidth]{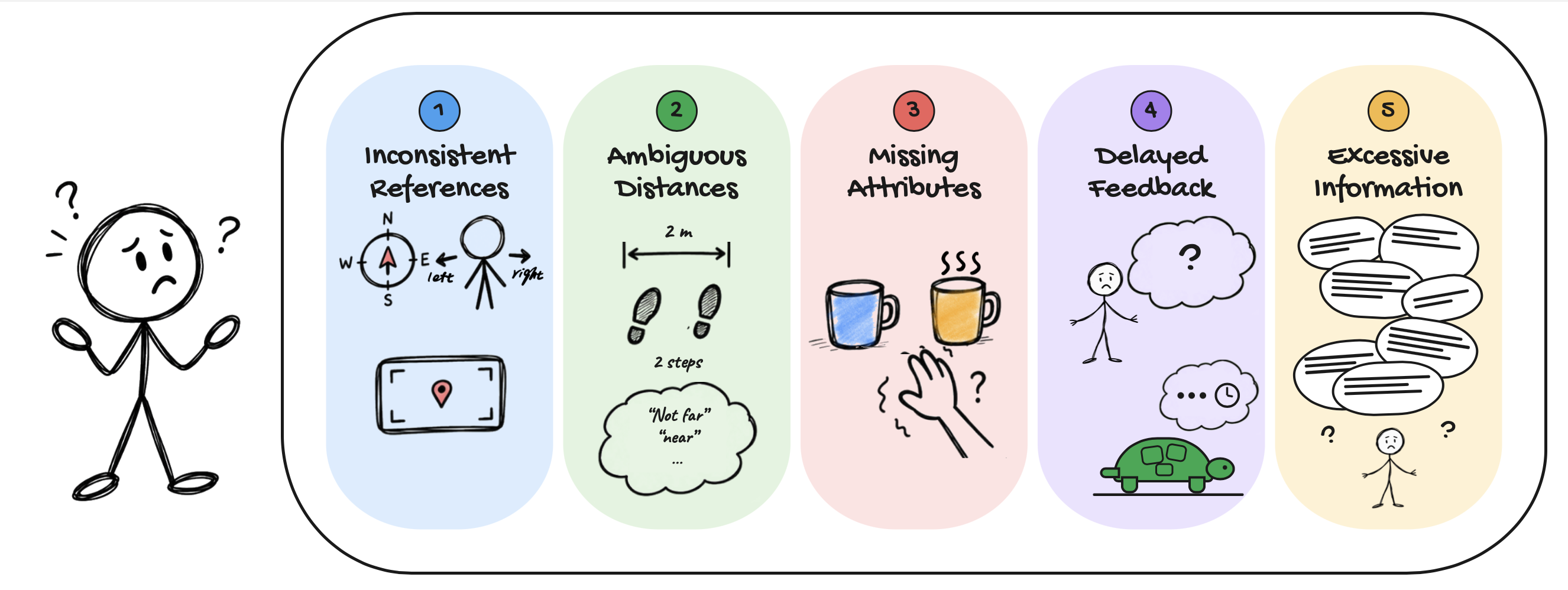}
  \caption{Five challenges with current LLM-powered visual assistance identified through our formative study: (1) inconsistent spatial reference frames, (2) distance descriptions that are difficult to translate into movement, (3) missing tactile attributes for distinguishing objects, (4) delayed feedback during movement, and (5) excessive information unrelated to the current action. }
  \label{fig:challenges}
\Description{
The figure summarizes five challenges with current LLM-powered visual assistance identified in the formative study. A confused stick figure appears on the left, while five numbered panels on the right illustrate the challenges. 
(1) Inconsistent References: a compass, left/right arrows around a person, and a camera frame with a location pin represent conflicting spatial reference frames, such as world-centered, body-relative, and camera-relative directions. 
(2) Ambiguous Distances: examples including 2 m, 2 steps, near, and not far show that distance descriptions may be inconsistent or difficult to translate into movement. 
(3) Missing Attributes: two visually distinguishable mugs and a reaching hand illustrate that visual descriptions may omit tactile properties needed to identify an object by touch. 
(4) Delayed Feedback: a person speaks to an AI and waits for a response, while a clock and turtle symbolize the latency of current LLMs and guidance that may arrive too slowly during movement. 
(5) Excessive Information: many overlapping speech bubbles surrounding a confused user illustrate descriptions that provide more information than is useful for the user's immediate action.
}
\end{figure*}

We used open coding and axial coding over the transcribed text. The analysis identified five common challenges with the current AI tools, illustrated in Figure~\ref{fig:challenges}.

\textbf{C1. Unstable and dynamic switching among reference points increases cognitive load.} Existing AI systems (e.g., Doubao and Be My Eyes) often switch unpredictably among scene- (FP1, FP2), object- (FP1, FP3), and camera-centered frames (FP6). For example, an object is described first relative to the room, then to another object, and then to the camera view. Participants could not reconcile these shifts the way sighted users do at a glance and had to mentally reconstruct the scene before acting (FP1, FP2, FP3, FP6). 

\textbf{C2. Existing metrics make instructions hard to execute.} Current tools often express directions in vision-oriented terms such as ``1--2 meters away'' (FP1, FP5) and ``slightly to the left front.'' (FP4, FP7) Such expressions carried little meaning participantas during movement as these metrics are not actionable nor visible for them. Participants believed instructions became more actionable when spatial relations were expressed in embodied units (FP2, FP5, FP8). For examples, step counts might map distance onto locomotion (FP1, FP7), while arm's reach and hand spans could express the remaining relation at reaching scale (FP2, FP7). 

\textbf{C3. Missing tactile properties prevent users from disambiguating similar objects.} Participants heavily relied on touch to distinguish among objects in unfamiliar environments, but current tools often missed properties of these objects. Even when participants knew the type of object they were looking for, they might not know the material, texture, size, or other physical features of the specific target. Useful properties were \textbf{material} of a cup, the \textbf{texture} of a cloth, or how a surface should \textbf{feel} (FP7, FP8,FP1, FP3, FP4, FP6). 

\textbf{C4. Delayed feedback makes spatial guidance stale during movement.} Participants reported noticeable delays when using current AI visual-assistance tools (FP1, FP2, FP3, FP4, FP8). This was particularly problematic during movement, because spatial relations changed continuously with their position and orientation (FP3, FP8). FP8 explains that an object described as being on the her right could be at the front after the user walked a few steps. Participants described having to stop, wait for an updated response, or query the system again before continuing (FP1, FP4). They wanted guidance to update with their movement so that directions remained aligned with their current pose (FP2, FP4, FP8).

\textbf{C5. Excessive information obscures what is relevant to the user's current action.} Participants reported that current tools sometimes provided too much information at once, including an object's color, shape, distance, direction, nearby objects, and broader scene context (FP2, FP3, FP5, FP7, FP8). Much of this information was unnecessary and could make the guidance harder to follow (FP5, FP7). Participants wanted information to match their immediate sensing and action needs. Some want concise directional or step-based guidance while moving (FP3, FP5), and richer tactile information when acquiring the target (FP2, FP5, FP7). 

The five challenges informed the iterative design of Touvigation described in Section~\ref{sec:system}.

\section{Touvigation System}

\label{sec:system}

Based on the findings, Touvigation aims to support not only finding where an object is, but also how to guide the BLV users actually acquire them. One main challenge remains the large body of literature is that \textbf{when a system identifies a distant object for the user, how can we continuously translate the goal into a series of actionable, embodied movement that helps to acquire the object.} A vision-based recognition is a starting point; as users move, target may shift in-and-out the FOV; as users approaching a distant object in a large indoor space, scales of actions change and the initial recommendations needs updates; right before reaching the object (e.g., within reach), the users need to move their upper body rather than full-body movement. 

Based on these design considerations, Touvigation transforms object search from a sequence of visual descriptions into a continuous embodied control process, maintaining a persistent spatial target while progressively changing the reference frame of guidance as the user's actionable space shifts from locomotion to actual acquiring. The goal of the system is to ensure that even if a tracked object is out of FOV, it will still be fully tracked and used to guide the BLV user back to the right track. The system is achieved via three main steps, the first step is to associate 2D visual detection with 3D entities and reconstruct the 3D environment for continues guidance; the second step is to close the BLV users' locomotion loop with adaptive body anchors, object avoidance, and route planning; and finally our system automatically re-anchors between body movement and hand movement at various interaction distances until the target is acquired. 
\begin{figure*}[t]
  \centering
  \includegraphics[width=0.85\linewidth]{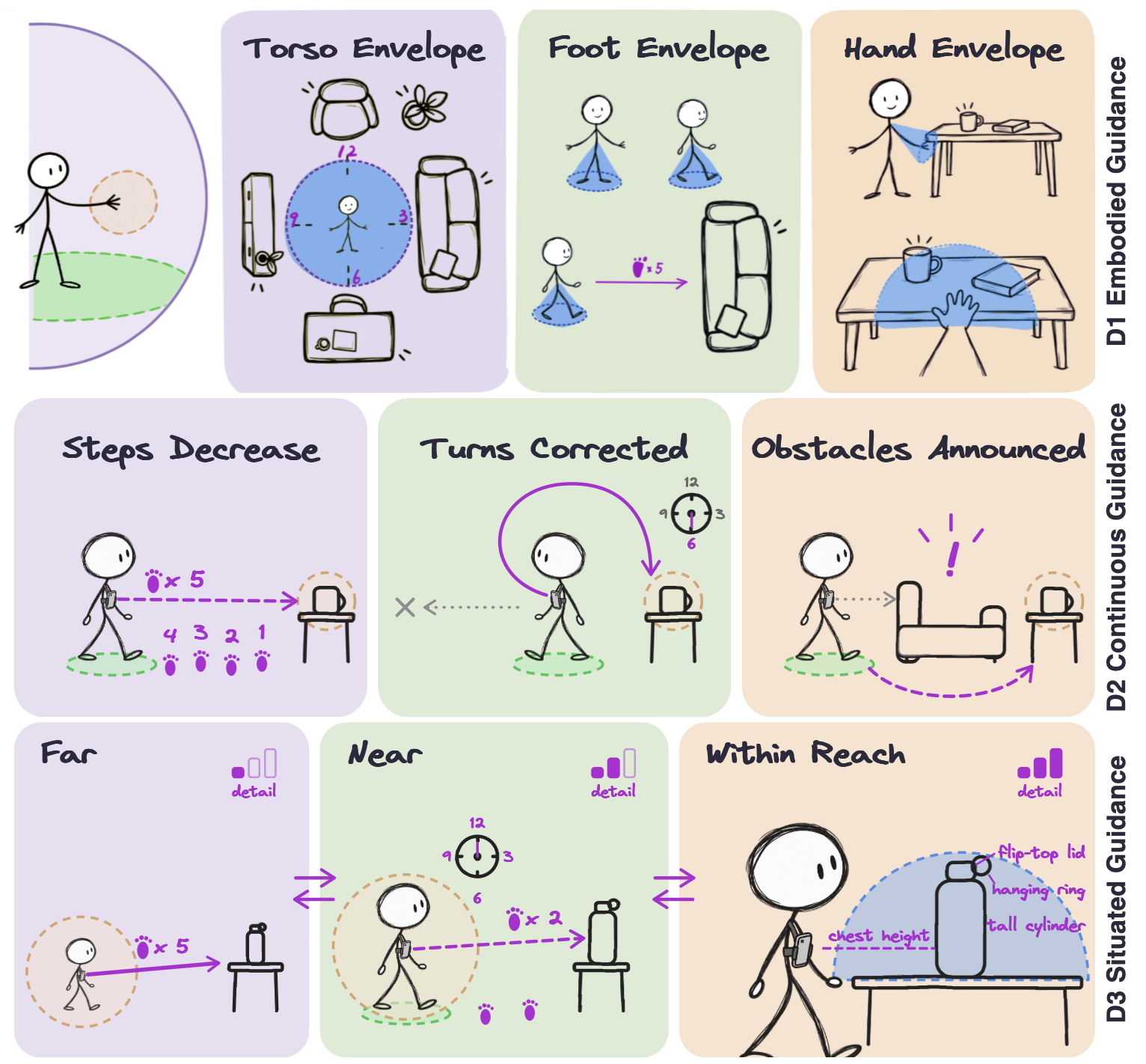}
  \caption{The design rationale of Touvigation. D1: embodied guidance uses torso, foot, and hand envelopes for direction, distance, and final reach. D2: continuous guidance updates instructions and corrects errors during movement. D3: situated guidance progressively increases detail as users approach the target.}
  \label{fig:envelopes}
\Description{
The figure presents three design principles of Touvigation in a hand-drawn, three-row illustration using lavender, green, and orange panels. Stick figures, arrows, footprints, clock faces, and colored regions visualize how guidance changes around the user's body and during movement.

D1, Embodied Guidance, uses three body-centered envelopes. The torso envelope places the user at the center of a clock face with furniture around them to show body-relative directions. The foot envelope uses walking direction and step counts, while the hand envelope shows a reaching region over nearby objects on a table.

D2, Continuous Guidance, shows guidance updating during movement. Step counts decrease as the user approaches the target, a wrong turn is corrected with a curved arrow and new clock direction, and an obstacle is marked with an exclamation point while a dashed path redirects the user around it.

D3, Situated Guidance, progresses from Far to Near to Within Reach. Detail increases from coarse step counts, to step counts plus clock directions, and finally to close-range spatial and tactile descriptions such as chest height, tall cylinder, hanging ring, and flip-top lid.
}
\end{figure*}
\subsection{Design Space}
The challenges in Section 3 motivate three design decisions: where and in what units guidance is expressed (D1), when and how often it is delivered (D2), and what information it conveys about the target (D3). 
\subsubsection{D1: Embodied Guidance.}
Guidance should be expressed in relation to the body. We organize guidance into three nested body-centered envelopes, each corresponding to the body part primarily involved in action at that scale (Figure~\ref{fig:envelopes}). The \textit{torso envelope} supports orientation by expressing direction relative to the user's current facing using clock bearings (e.g., ``at your 2 o'clock''). The \textit{foot envelope} supports locomotion by expressing distance in steps rather than meters. The \textit{hand envelope} supports acquisition by expressing body-relative height, palm-scale movements, and tactile properties such as material, shape, and texture. We introduce a common embodied vocabulary for describing where to orient, how to move, and how to reach and identify an object.
 
\subsubsection{D2: Continuous guidance}
Guidance should operate as a closed loop between the user's movement and the system's feedback. During navigation, the relevance of an instruction changes as soon as the user moves. A direction can become incorrect, a remaining distance can shrink, or a new obstacle can enter the path. Touvigation should maintain a persistent 3D target and continuously relates it to the user's current pose. Wrong turns and out-of-view objects should be corrected before they accumulate. Nearby obstacles should be announced when they fall on the path. 

\subsubsection{D3: Situated guidance.}
Guidance should dynamically adapt to how the user is currently sensing and progressing toward the target. Our system should limit the type of information conveyed based on the user's current goal and interaction state to avoid information overflow. For instance, coarser directional information can be used at far away orientation and locomotion, but finer spatial, shape, material, and texture information can fill in instructions when objects are within reach, see Fig~\ref{fig:envelopes}. These envelopes are not fixed stages in a one-way sequence. If the user explores the expected location by hand but does not find the object, the system can exit the hand envelope and return to foot- or torso-level guidance to support renewed localization and movement. Guidance thus moves dynamically between envelopes as the user's sensing strategy and task state change.

\begin{figure*}[t]
  \centering
  \includegraphics[width=\textwidth]{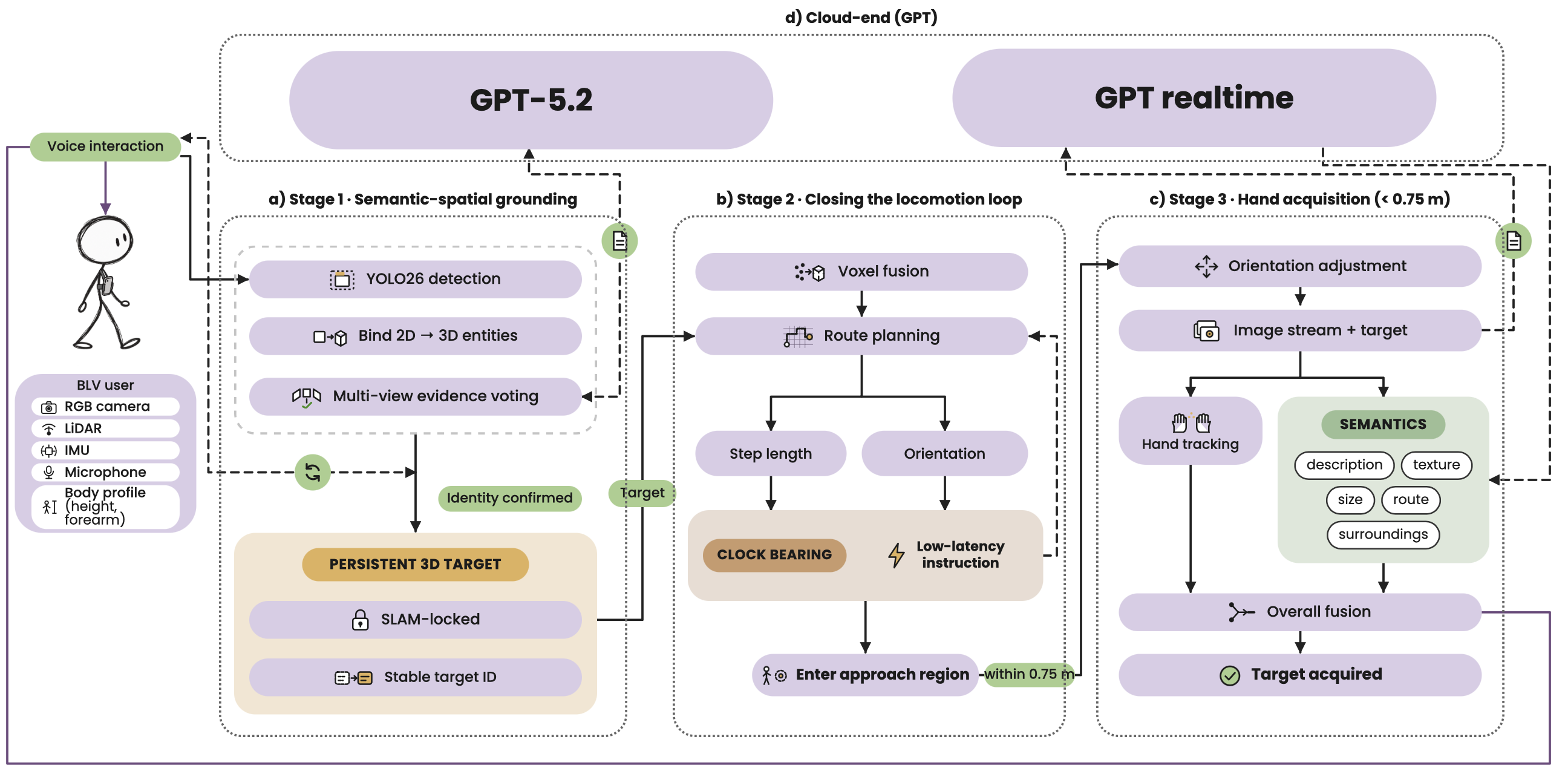}
  \caption{System overview of Touvigation. (a) Semantic-spatial grounding binds visual detections to a persistent 3D target. (b) Continuous locomotion guidance updates the route using clock directions and personalized step counts. (c) Hand acquisition combines target semantics with hand tracking to guide reaching and confirmation. (d) Cloud models support object verification and semantic descriptions, with guidance delivered through voice.}
  \label{fig:system-flow}
  \Description{A flowchart. On the left, a BLV user with RGB camera, LiDAR, IMU, microphone, and a body profile provides input and receives voice guidance. Stage 1 turns YOLO26 detections ~\cite{jocher2026ultralytics} into a persistent SLAM-locked 3D target through 2D-to-3D binding and multi-view evidence voting. Stage 2 runs voxel fusion and route planning, producing clock-bearing and low-latency instructions using step length and orientation, until the user is within 0.75 meters. Stage 3 adjusts orientation, sends the image stream and target to the cloud for semantics, and fuses the result with hand tracking until the target is acquired. A cloud section at the top holds GPT-5.2 and GPT realtime.}
\end{figure*}

\subsection{Persistent Semantic-Spatial Grounding}
One of the main challenge for the system is how to integrate instant physical object label identification into persistent embodied navigation. Our preliminary testing with YOLO26 on a Iphone 17pro in the lab space found that YOLO26 alone failed to provide consistent object labeling even if the object did not move; The the identification start to collapse and mistaken the same object for other labels when the user starts to move. This will make the system's navigation fail as it relies on the previous frame as reference. As a result, we propose an integration strategy to convert 2D sparse YOLO26 detection into persistent 3D entities using Simultaneous Localization and Mapping (SLAM), commonly used for AR world tracking and user localization. The strategy first use YOLO26 to get bounding boxes of physical objects and its labels, and then binding the label to SLAM world's 3D object after a voting algorithm verifies the label with an online LLM (ChatGPT 5.2). The 3D object will then be used to render the object's label, support navigation, and provide references, see Fig. ~\ref{fig:system-flow} for the complete system flow. 

\subsubsection{Associating Visual Detections with 3D Entities}
To infer \textit{what is the label of a target physical object}, we map the YOLO26's 2D bounding box to the 3D object. This is achieved by using the 2D bounding boxes information from YOLO26 and check they intersections with the projections of target 3D objects' bounding boxes on the screen space with an AABB algorithm. To further disambiguate target 3D objects from others, we use ARKit's depth information ~\cite{apple2026arkit} to provide each 3D object an estimated distance, temporary ID, and estimated dimension for later LLM use. However, handling situations where two similar small size that occludes each other (a vase occludes a napkin nearby box) remains future work. As a result, we capture a series of raw images using the intersection of bounding boxes in the video stream, labels of the target objects along with other identified nearby objects, and physical properties of them. The system then proceed with a multi-view evidence voting to find the most likely object using LLM's semantic reasoning, disambiguate occlusions, and updates the temporary ID of the target object. We keep the retained views separate rather than compositing them into a single panorama~\cite{opencv2026stitcher}, and we avoid general-purpose segmentation backbones~\cite{kirillov2023segment}, which are designed for desktop-class hardware, so that the on-device stage stays within the phone's real-time budget.

\subsubsection{Multi-view evidence voting}. Essentially the voting is a Bayesian-Inspired synthesis. The user's phone provides spatial computation and enforce the LLM to perform reasoning under constrains from the depth and point cloud data from the smartphone. We describe the estimation in three layers. 

\textbf{Layer 1: Spatial and view quantization} We combine a series of images from the AR video stream using the 60 degree spatial azimuth hash to bin the observation viewing angles. A Top-K algo is used to filter key frames using multivariate fitness equation $q$ for maximum gain. 
\begin{equation} q = w_1 S_{\mathrm{det}} + w_2 \min\left(1,\frac{A}{A_0}\right) + w_3 R_{\mathrm{depth}} + w_4 C - w_5 \frac{\lvert \Delta P \rvert}{\Delta t}, \end{equation}
where ($S_{\mathrm{det}}$) denotes the confidence score produced by the local YOLO26 detector. The term ($A/A_0$) represents the normalized bounding-box area. ($R_{\mathrm{depth}}$) measures depth reliability (C) denotes the image-center proximity score. Finally, ($\lvert \Delta P \rvert / \Delta t$) serves as a motion penalty, using the camera angular velocity as a physical proxy for motion blur. The weights ($w_1,\ldots,w_5$) control the relative contribution of each term.

\textbf{Layer 2: Depth-aware LLM inference}
To mitigate the scale ambiguity from LLMs, we implement a state-constrained prompting procedure that seperates visual interpretation from geometry-aware reasoning. The prompt includes images and absolute geometric piriors (i.e., objects bounding box, depth, point clouds) from ARKit including estimated physical size ($S_{\mathrm{real}}$) and its  height($H_{\mathrm{real}}$) from the ground. The LLM is instructed to reconcile its first visual hypothesis with the provided physical measurements before providing the final semantic label. This helps the LLM to ``sense-making'' an object especially as they are likely to be low-pixel objects that only occlupies a small portion of the scene (e.g. distant objects). For example a coffee cup over 30 centimeters wide is as unlikely as a front-door of 20 centimeter. The final prediction can be expressed as \begin{equation} L^{*} = \arg\max_{L} P(L) P(I\mid L) P(S_{\mathrm{real}}\mid L) P(H_{\mathrm{real}}\mid L), \end{equation} 

where ($I$) denotes the visual observation and (L) the candidate semantic label. 

\textbf{Layer 3: Synthesis and Fusion Gate.} 
After LLM returns, the smartphone aggregates three sources of semantic evidence into a final score ($S(L)$) for each candidate label ($L$). The fusion assigns the largest contribution to multi-view consensus, while the local detector prior and the LLM inference provide complementary evidence:

\begin{equation}
S(L)
=
\alpha P(L\mid \mathrm{YOLO26})
+
\beta
\frac{
\sum_{v=1}^{K}
q_v c_v
\mathbb{I}(n_v=L)
}{
\sum_{v=1}^{K} q_v
}
+
\gamma P(L\mid \mathrm{LLM}),
\end{equation}

where ($K$) denotes the number of retained views, ($q_v$) is the quality score of view ($v$), ($c_v$) is its semantic confidence, and ($n_v$) is the predicted label. ($\mathbb{I}(\cdot)$) is the indicator function, while ($\alpha$), ($\beta$), and ($\gamma$) control the relative contributions of local detection, multi-view agreement, and LLM inference. We set $\alpha$ to be 0.1, $\beta$ as 0.25 and $\gamma$ as 0.65 to maximize LLM's influence and importance of multiple image sequences (Multiview). Gating is applied to the $S(L)$ to determine whether the label preserves or overrides. 

\subsubsection{Persistent Target Locking}
Once target's identity is confirmed, our system begins the 3D entity tracking in the physical environment. The system records the target's 3D bounding box within the SLAM world. This way, when the user moves, rotates in the physical space, the 3D bounding box moves and rotates along. Additionally, 
This design allows users to temporarily exit the camera's field of view without losing the navigation target, as SLAM tracking provides quick and accurate relocalization once the same scene is back to the view. No further 2D visual detector to identify the target is required in each frame, enabling a 10hz, low-latency tracking experience.  

\subsection{Closing the Locomotion Loop with Adaptive Body Anchors}
Our system aims to deliver an embodied target acquiring experience beyond traditional route planning from point A to B. To achieve that, users at various distances must receive suitable instructions to best fit their body movements. During a locomotion stage, we use users' body as the main spatial anchor, converting navigating path into personalized step counts with step length estimation. 

\subsubsection{Geometric Target Planning}
ARKit provides sparse point clouds and directly use it as a navigation surface is unstable. The system then applies voxel fusion results to represent indoor geometry for navigation, extracting the floor plane and identifying impassable areas. The walkable space is then converted into a two-dimensional occupancy representation, and A* search ~\cite{hart1968formal} is employed to determine a collision-free route from the user's current location to the vicinity of the target.

To optimize the raw A* path, The system simplifies it using a string-pulling technique to eliminate unnecessary local vertices. Navigation instructions are updated based solely on the current straight-line segment and the next critical turn, rather than presenting the entire planned path to the user simultaneously. The objective of this navigation phase is not reaching the exact center of the target object, but rather to arrive at an approach region suitable for initiating the subsequent reaching interaction. Therefore, the endpoint of the path planning process is defined as a body position that enables a safe transition from walking to reaching.

\subsubsection{Personalizing Steps}
Touvigation translates long-range geometric distances into an \textit{adaptive step space}.To calculate this translation, our system takes a current path segment, let $\mathbf p_t$ denote the user's current position and $\mathbf g$ represent the endpoint of the current straight-line segment. The remaining two-dimensional distance is given by $
d_t = \left\| (g - p_t)_{xz} \right\|
$. The system then quantizes the distance with the estimated step size $\hat{s}_t$ as 
$N_t = \max\left(1,\operatorname{round}\left(\frac{d_t}{\hat{s}_t}\right)\right).$

The current implementation updates this state approximately every six AR frames, or about $10hz$. The step detection directly triggers the announcements, and users will hear how many steps left to reach the current segment (i.e., before a turning instruction) in realtime. Spoken requests and spoken guidance are handled through streaming speech-to-text and text-to-speech services~\cite{openai2026stt,openai2026tts}, and the conversational component runs on a streaming realtime interface so that a reply can begin before the full response is generated~\cite{openai2026realtime}. 

Touvigation's step calculation is personalized to fit users' distinct stride. To do so, the system uses  cumulative step counts measured by CMPedometer with the actual displacement obtained from ARKit Visual-Inertial Odometry (VIO) $ s_{\mathrm{obs}} = \frac{D_{\mathrm{VIO}}}{N_{\mathrm{pedo}}}.$ The step length is then updated incrementally using an Exponential Moving Average (EMA). 
Consequently, the same 3 meters distance may correspond to different numbers of physical movements depending on the user. The term ``step'' is not a fixed linguistic unit but a personalized spatial metric.

\subsubsection{Continuous Body-Centric Guidance}
The system defines the angle between the user's current horizontal orientation and the subsequent path segment as the signed angle. This angle is then assigned to a specific ``clock direction'' schematic, such as 12 o'clock, 1 o'clock, 3 o'clock, or 9 o'clock using $30^\circ$ intervals. Angles that are close to the forward direction and minor body swing do not trigger reorientation. The implementation considers deviations smaller than approximately $15^\circ$ as straight ahead.
Therefore, a path segment may be represented as \textit{Turn left to the 9 o'clock position, then proceed forward for three steps.}

This is not a one-time route description. The system continuously reads the user's latest camera pose and calculates the new straight-ahead segment and the remaining steps. When the user moves in the correct direction, the remaining step count decreases. If the user deviates, overshoots, or leaves the planned route, the system will respond with in 100ms and update according to the new physical location instead of following outdated instructions.

\subsection{Re-Anchoring Within Reach}
Merely increasing navigation precision is insufficient to acquire a target. A core issue is that as the user approaches the target, the initial spatial reference frame loses its utility. When the target is several meters away, the user’s primary actions are turning and walking; however, when the target is only tens of centimeters in front of the user, whole-body instructions may cause over-movement. Therefore, Touvigation facilitates a \textbf{reference-frame transition}, gradually shifting the spatial anchor from the whole body to the arm and hand.

The system transitions to the \textit{hand-guided stage} when the user is approximately $0.75$ m within the target. The stage transition is implemented as a state process. To avoid rapid toggling between locomotion and reaching modes when the user moves near the boundary,  we use hysteresis with a larger exit distance and a minimum duration requirement.

At the onset of the near-target stage, the system initiates a local re-orientation process to ensure the user to perform fine-grained lateral adjustments to align their body and device with the target.

\subsubsection{Hand-Guided Acquisition}
Once the re-orientation is ready, the system generates a local scene snapshot, sending over to the MLLM that returns whether the target sits on a larger object and whether nearby semantic objects are present. Additionally, a touch route is computed to describe how to use hand touch to reach the final object. The objective is not to compute an exact robotic trajectory, but to identify stable reference points that enable the user to progressively narrow the search area through tactile exploration. Here we implement a large-to-small priority prompt template that forces MLLM to describe the touch route from larger, nearby objects to smaller ones. For example, if the target is a cup on a table, the system may recommend locating the table edge or surface first, then describing the target’s position relative to this large-scale tactile anchor. Similarly, if the target is on a shelf, the system may guide the user to the shelf structure or a specific shelf board before directing attention to the smaller target. Nearby objects function as pre-touch disambiguation context; by informing the user of other candidate objects near the target, the system facilitates the construction of a local mental map.

\subsection{3D Hand Tracking for Guiding Hand Guidance}
Upon the touch route is ready, Touvigation uses the \textit{VNDetectHumanHandPoseRequest} from Apple's Vision framework to detect a maximum of two hands. The system designates the index fingertip as the primary hand proxy and defaults to the wrist when detection fails. Hand detection is performed every 6 frames, resulting in an approximate rate of 10 Hz.

Because the Vision framework provides raw 2D joint coordinates, our system incorporates LiDAR depth data to reconstruct their 3D positions. For each joint pixel, the system samples valid depth values using a $3\times3$ medium kernel to reduce depth noise. Camera's intrinsics and the camera-to-world transformation are applied to obtain the hand's world coordinates ($\mathbf h_t$). The same pipeline is not tied to one platform, since comparable on-device hand and body pose estimators exist elsewhere~\cite{google2026mlkitpose}; it does, however, depend on a device with LiDAR, and monocular metric depth estimators~\cite{yang2024depthanythingv2} are a plausible substitute on phones without one.

\subsubsection{Calculating Body-centric Hand Guidance}
We let $B$ as the 3D label of the target object obtained in earlier voting, Given $\mathbf h_t$, the system first computes the closest point:
$\mathbf p_t^* = \operatorname{ClosestPoint}(\mathbf h_t, B).$
The corresponding hand-to-target displacement is
$\mathbf o_t = \mathbf p_t^* - \mathbf h_t.$ Instead of expressing this displacement in the global coordinate system, Touvigation transforms $\mathbf o_t$ into a body-centric coordinate frame defined by the user's current forward, right, and vertical axes:

\[
\mathbf o_t^{B}
=
\left(
o_{\mathrm{right}},
o_{\mathrm{up}},
o_{\mathrm{forward}}
\right).
\]

This transformation converts a geometric displacement in 3D space into directly executable reaching actions. For example, a positive lateral offset is translated as ``move your hand to the right.''

At this stage, the process no longer depends on the target's visibility in the camera view. Since the target's semantic identity and 3D location have already been persistently grounded, the hand-guidance loop only updates the user's hand position relative to the fixed target geometry. This enables the system to maintain continuous reaching guidance even when the user's hand partially occludes the target or the target temporarily leaves the camera image.

\subsubsection{Hand-scale Units}
While the scale during the locomotion stage was ``steps'', our system now uses forearm and palm size as units. For distances less than users' forearm the system automatics reports in number of palms. At this scale, the orientation also uses ``left'' and ``right'' instead of the clock-wise design. The forearm length is defined as $0.146 * BodyHeight$. As a result, distances are reported as ``one forearm in front of..'' or ``two palms on the left side''.

Touvigation maps the target height to body-relative vertical zones, such as ankle, knee, waist, chest, and shoulder. It then translates these zones into postural cues, such as ``squat down,'' ``bend over to touch,'' or ``reach up.'' Additionally, scale the body-height zones proportionally according to the user's height.

\subsubsection{Voice confirmation for object acquisition}
We define less than 10 cm between the users' palm center to the object's outer bounding box as \textit{object reaching} state. The 10cm threshold is empirically setup to match the hand tracking accuracy for disambiguiation. Although a BLV user can already feel the object as they touch the target, it is still beneficial for the system to confirm since similar feeling object could be mistaken for other geometrically similar objects. 

\subsection{Balancing Responsiveness and Feedback Stability}
Our system categorizes computational tasks according to how quickly an error becomes actionable. The system operates primarily across two layers: the semantic loop, which manages object descriptions and functionality, and the motor loop, which processes motion and tracking information with low latency. For example, during hand acquisition, once the semantic loop establishes the touch route, it terminates further computation to ensure rapid response.

However, low latency does not imply that feedback should be delivered at an unlimited frequency. If the system were to provide immediate verbal feedback for every minor pose shift, such as a few centimeters, sensor noise or natural body sway could cause the audio output to rapidly alternate between “left” and “right.” To address this, Touvigation separates the state update rate from the speech update rate.

We measure the time it takes for different components in our system. This is achieved through a pilot study over 50 trials for locating different objects in our lab. On average, YOLOv8 took about 30ms to process its entire pipeline. During the object-searching stage (before voting), our system takes 3 snapshots over 1 second and sends them to the MLLM, with an average response time of 3.2 seconds. Once it finds the object, the system tracks it at ARKit’s native framerate of 60 FPS.  Hand detection took an average of 20 ms to form 21 joint points with reverse projection mapping.



            



    
    

\section{User Study}
\label{sec:userstudy}

We conducted a within-subjects study with 12 blind participants performing an object-finding (seek-touch-and-confirm) task in two unfamiliar rooms under three conditions: Touvigation, Doubao~\cite{bytedance2026doubao}, and unassisted search. Doubao is a leading commercial multimodal voice assistant from ByteDance, and it was the AI aid every participant in our sample already used, so it gives us a baseline that is both a strong contemporary VLM system and an ecologically valid reflection of how blind users find objects with AI help today. In addition, the effect of metric guidance was not directly measured due to findings from earlier formative study and literature, as the metric guidance are known to be non-intuitive for BLV users.

The study addressed three research questions:
\begin{itemize}
    \item \textbf{RQ1:} Does Touvigation improve the completion time and success rate of object finding for blind users in unfamiliar indoor environments, compared with a state-of-the-art AI assistant?
    \item \textbf{RQ2:} Does Touvigation reduce cognitive workload during object finding?
    \item \textbf{RQ3:} Does Touvigation improve users' perceived safety, trust, and spatial awareness during object finding?
\end{itemize}

\subsection{Participants}
\begin{figure}[t]
  \centering
  \includegraphics[width=\linewidth]{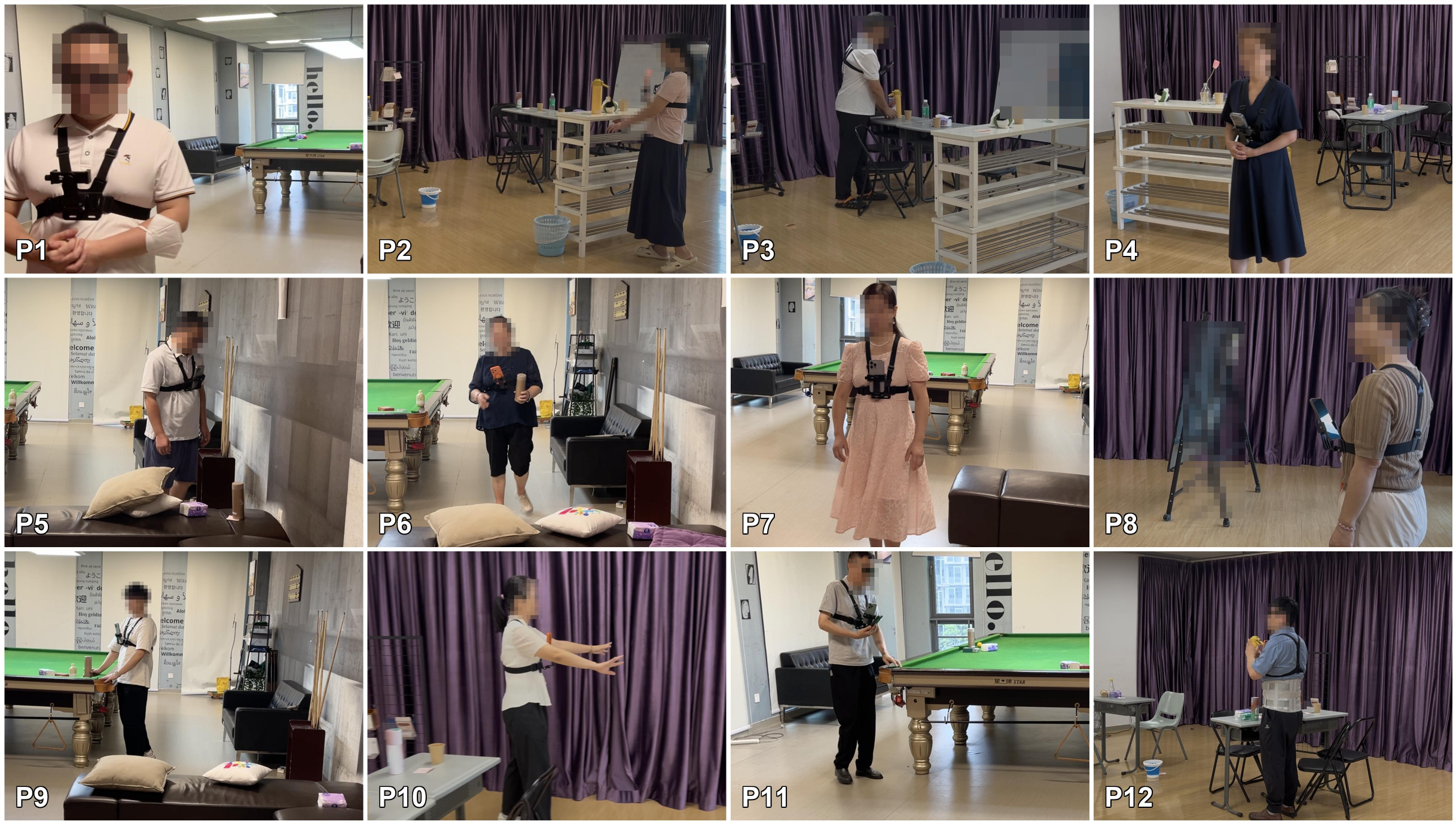}
  \caption{Participants P1--P12 during the study, each wearing the chest-mounted phone in one of the two rooms. Faces and selected content are pixelated for privacy.}
  \label{fig:participants}
  \Description{A four-by-three grid of photographs labeled P1 to P12. Each shows one participant standing or walking in the conference room or the activity room with a phone mounted on a chest harness; faces and any whiteboard writing are pixelated.}
\end{figure}

\looseness=-1 Twelve blind adults took part in the main study (6 women, 6 men; ages 23--55, $M=48.2$, $SD=8.8$). All twelve were blind at Level~1, the most severe of the graded categories under the Chinese national visual-disability classification standard, corresponding to best-corrected acuity in the better eye below roughly 0.02 or a visual-field radius under about 5$^\circ$~\cite{gbt26341}. Residual light perception and onset varied (Table~\ref{tab:participants}). All participants had prior experience with Doubao and other AI-based visual assistance. Several also used mainstream navigation apps and other assistive tools (Table~\ref{tab:participants}). All guidance in the study was auditory, and all participants had functional hearing. Figure~\ref{fig:participants} shows each participant during the study.

Participants were recruited through a partner at a local association for blind and low-vision people. 
Each received 150\,RMB (about US\$21) for the 1.5 to 2 hour session. The study was approved by our institution's Institutional Review Board (IRB).  Participants provided informed consent to the study with audio and video recording at the beginning of the session. Consent forms, interview questions and questionnaire items were read aloud and answered orally. Before the formal study, two pilot participants completed the protocol to provide first-round feedback; they are excluded from all analyses.

\begin{table}[t]
  \caption{Main-study participants (P1--P12). All were blind at Level~1 of the Chinese national visual-disability classification (see text). Residual perception: None = no light perception; Light = light perception only; ``+ color'' / ``+ slight color'' = additional partial color discrimination. Shike ~\cite{shike2026navigation} is a Chinese navigation app for blind users; Tencent, Baidu, and Gaode Maps are mainstream Chinese map apps.}
  \label{tab:participants}
  \footnotesize
  \setlength{\tabcolsep}{4pt}
    \begin{tabular}{@{}ll p{2.1cm} p{2.1cm} p{1.7cm} p{3.1cm}@{}}
      \toprule
      ID & Gender & Age & Residual perception & Onset & AI/assistive tools used \\
      \midrule
      P1  & M & 43 & None                 & Congenital    & Doubao, Shike \\
      P2  & F & 49 & Light                & Acquired (25) & Doubao, VoiceOver, Shike \\
      P3  & M & 52 & Light + color        & Acquired (16) & Doubao \\
      P4  & F & 53 & Light                & Congenital    & Doubao \\
      P5  & M & 46 & Light + slight color & Acquired (30) & Doubao, Be My Eyes \\
      P6  & F & 55 & Light                & Congenital    & Doubao \\
      P7  & F & 55 & None                 & Acquired (4)  & Doubao \\
      P8  & F & 55 & None                 & Congenital    & Doubao, Be My Eyes \\
      P9  & M & 23 & Light                & Congenital    & Doubao, Tencent Maps \\
      P10 & F & 47 & Light                & Acquired (40) & Doubao, Tencent Maps \\
      P11 & M & 48 & None                 & Congenital    & Doubao, Baidu/Gaode Maps \\
      P12 & M & 52 & None                 & Congenital    & Doubao, obstacle-avoidance wearable \\
      \bottomrule
\end{tabular}
\end{table}

\subsection{Conditions}
All three conditions were conducted in Mandarin. In C1 and C2 the assistant was chest-mounted on a LiDAR-equipped iPhone, an iPhone 15 Pro Max in one room and an iPhone 17 Pro in the other, since two participants ran in parallel.

\textbf{C1: Touvigation.} Participants used the full system described in Section~\ref{sec:system}, running on an iPhone worn in a chest harness with the rear cameras facing forward and both hands free.

\textbf{C2: Doubao.} Doubao was running on the same phone and was chest-mounted as in C1. Participants asked Doubao spoken questions turn by turn. We used Doubao as deployed at the time of the study (July 2026, version 13.4.0).

\textbf{C3: Unassisted.} No AI assistance. Participants were given the approximate location of the target but were not informed of its specific direction or height. They searched using their own strategies.

In every condition, the participant additionally wore a neck-mounted device running the data-collection app, which logged the 20\,Hz trajectory for every trial, so that motion data were captured consistently across study. All conditions ran on the same campus network, which provided connectivity for the Doubao app and for OpenAI API.

\subsection{Task and Environments}
In each trial, the participant began from a fixed start position, and had to locate and touch the thermos cup. The trial ended when the participant reached the object and confirmed finding personally. Only in the unassisted condition (C3) were they additionally given a coarse placement hint (e.g., ``on a table'' or ``beside the chair''). The hint was provided because an unassisted search could take arbitrarily long without any prior information. Distractor objects were present on the surfaces, so a touch counted as correct only when the participant identified the thermos cup to us. 

\begin{figure}[t]
  \centering
  \begin{minipage}{0.42\linewidth}\centering
    \includegraphics[width=\linewidth]{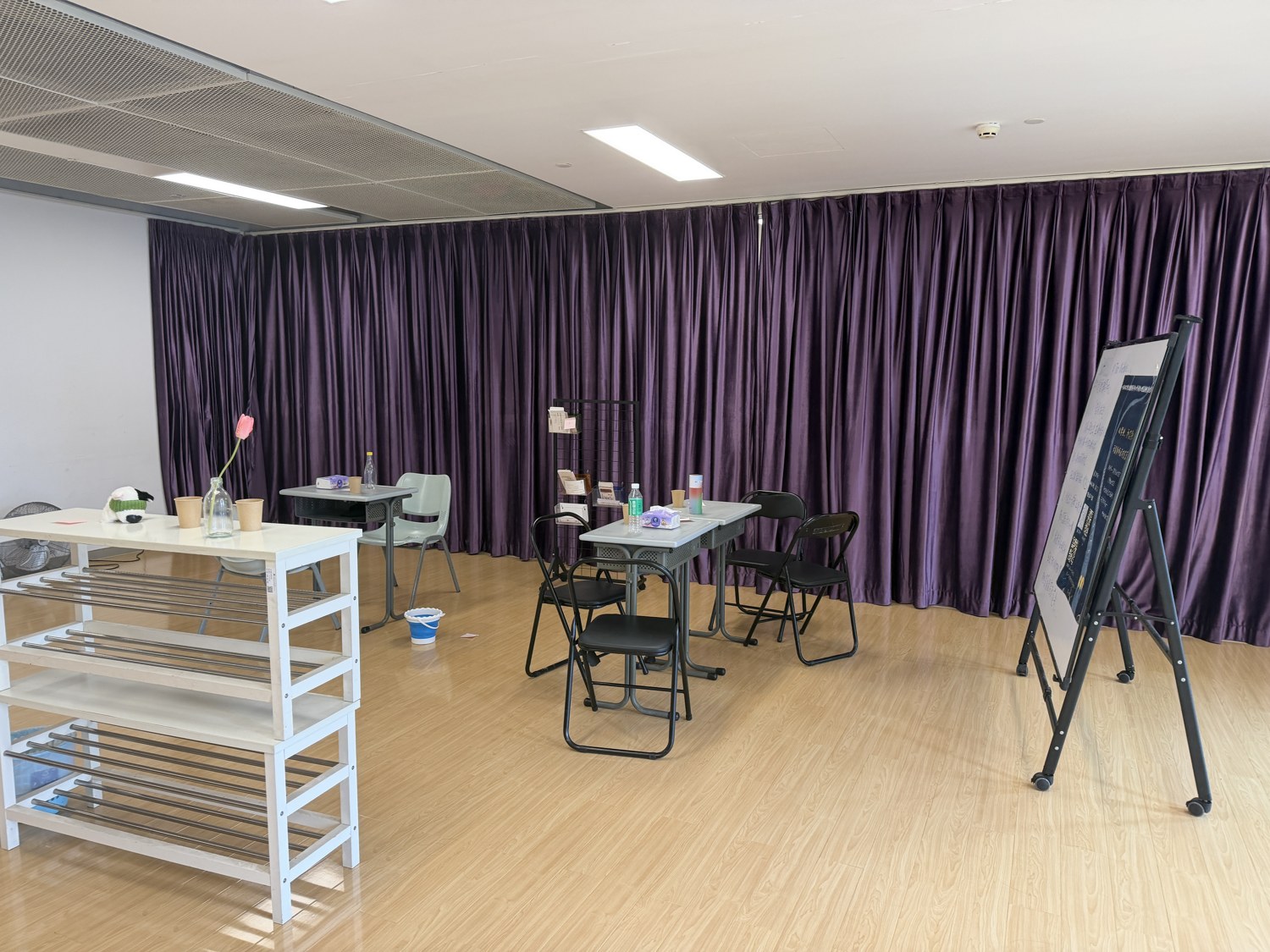}\\
    {\small (a) L1: conference room}
  \end{minipage}\hfill
  \begin{minipage}{0.42\linewidth}\centering
    \includegraphics[width=\linewidth]{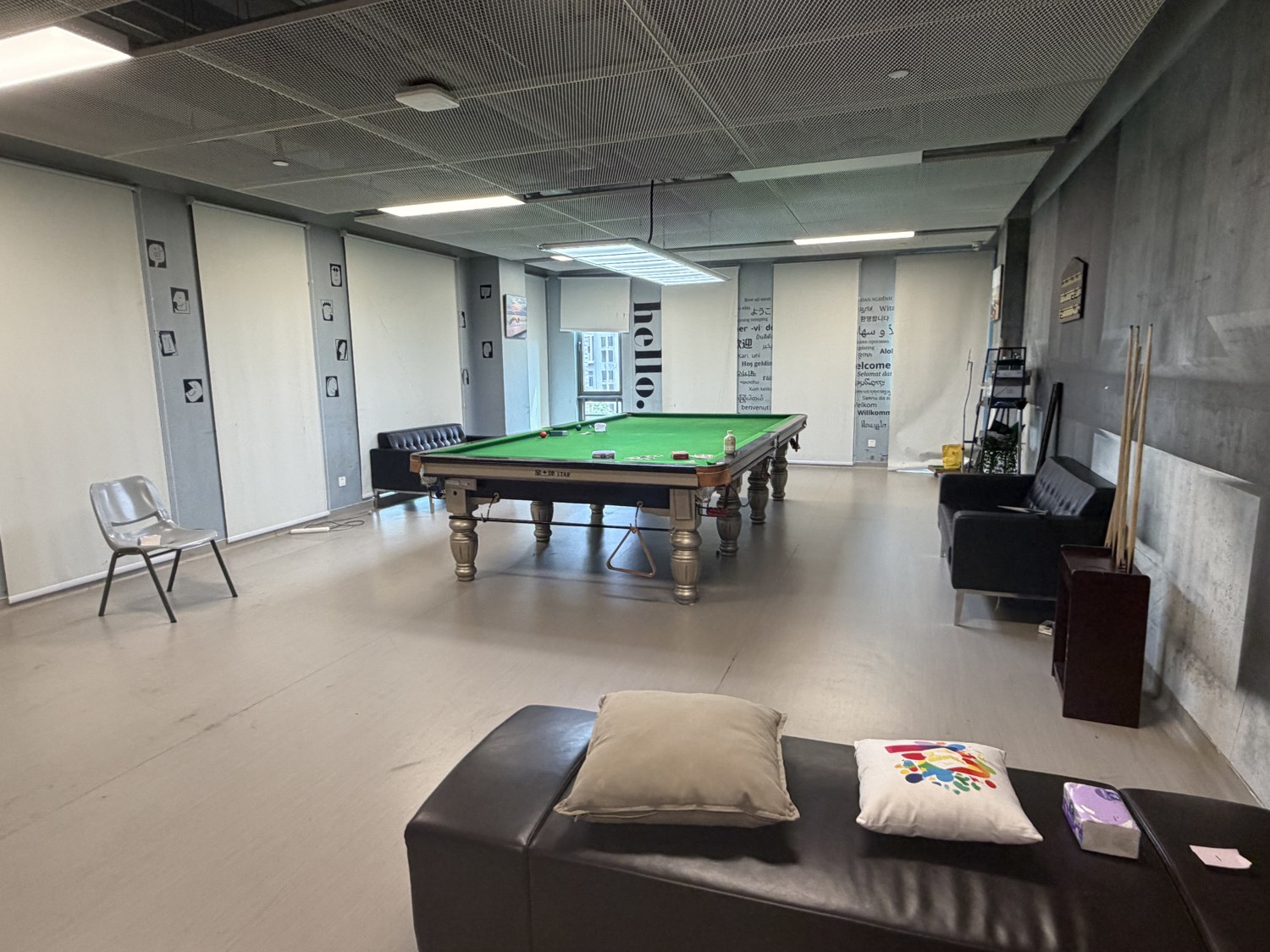}\\
    {\small (b) L2: activity room}
  \end{minipage}

  \vspace{4pt}
  {\centering\footnotesize\textbf{L1 target positions}\par}
  \vspace{1pt}
  \begin{minipage}{0.155\linewidth}\centering\includegraphics[width=\linewidth]{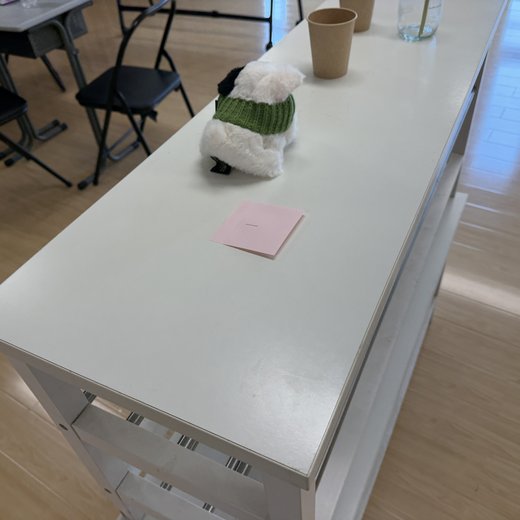}\\{\scriptsize Pos 1}\end{minipage}\hspace{1.5pt}%
  \begin{minipage}{0.155\linewidth}\centering\includegraphics[width=\linewidth]{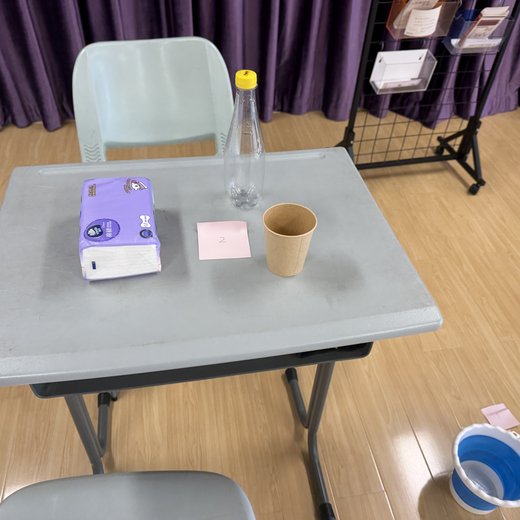}\\{\scriptsize Pos 2}\end{minipage}\hspace{1.5pt}%
  \begin{minipage}{0.155\linewidth}\centering\includegraphics[width=\linewidth]{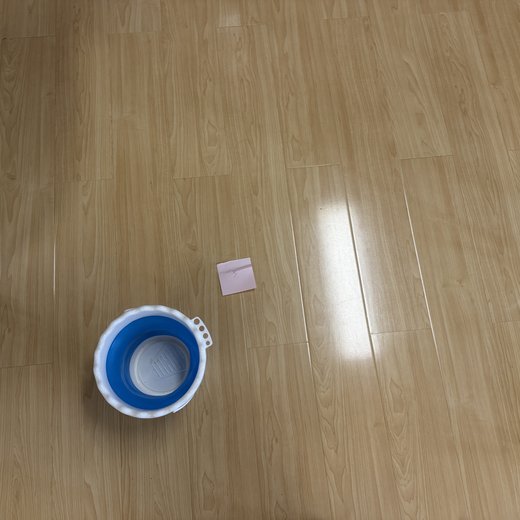}\\{\scriptsize Pos 3}\end{minipage}\hspace{1.5pt}%
  \begin{minipage}{0.155\linewidth}\centering\includegraphics[width=\linewidth]{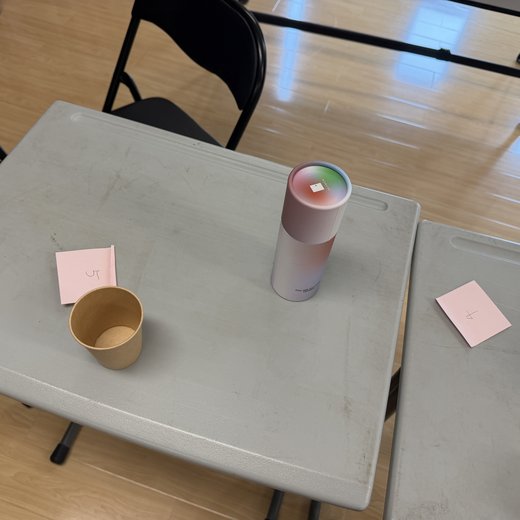}\\{\scriptsize Pos 4/5}\end{minipage}\hspace{1.5pt}%
  \begin{minipage}{0.155\linewidth}\centering\includegraphics[width=\linewidth]{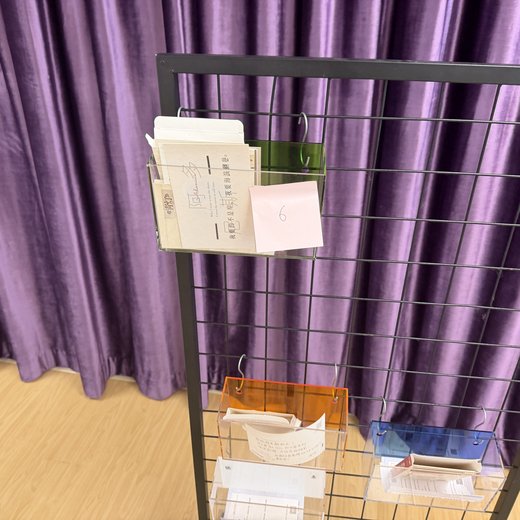}\\{\scriptsize Pos 6}\end{minipage}

  \vspace{4pt}
  {\centering\footnotesize\textbf{L2 target positions}\par}
  \vspace{1pt}
  \begin{minipage}{0.155\linewidth}\centering\includegraphics[width=\linewidth]{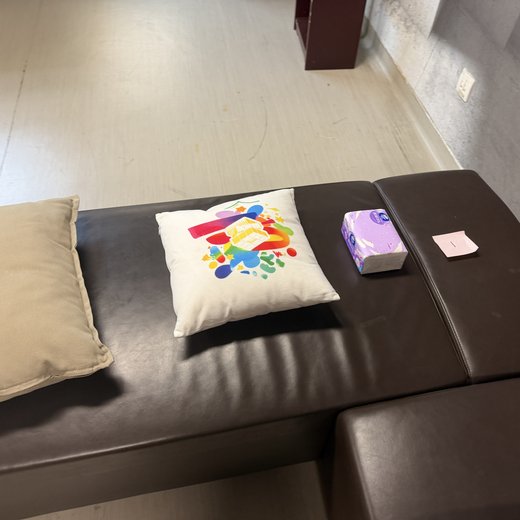}\\{\scriptsize Pos 1}\end{minipage}\hspace{1.5pt}%
  \begin{minipage}{0.155\linewidth}\centering\includegraphics[width=\linewidth]{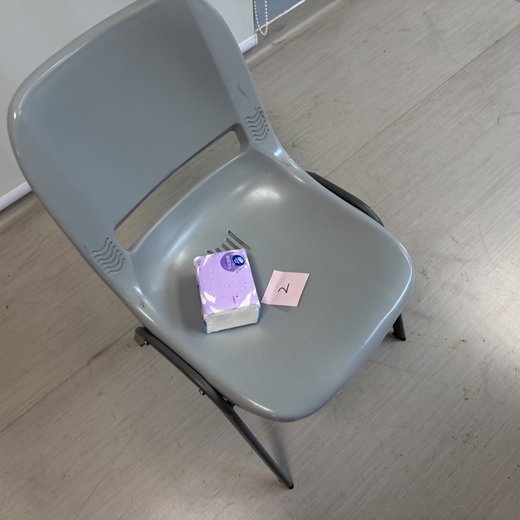}\\{\scriptsize Pos 2}\end{minipage}\hspace{1.5pt}%
  \begin{minipage}{0.155\linewidth}\centering\includegraphics[width=\linewidth]{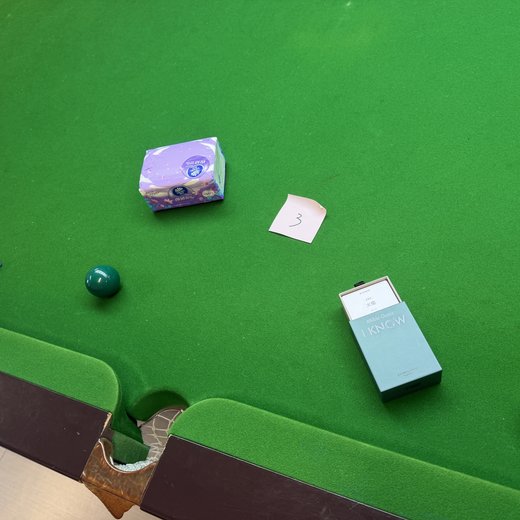}\\{\scriptsize Pos 3}\end{minipage}\hspace{1.5pt}%
  \begin{minipage}{0.155\linewidth}\centering\includegraphics[width=\linewidth]{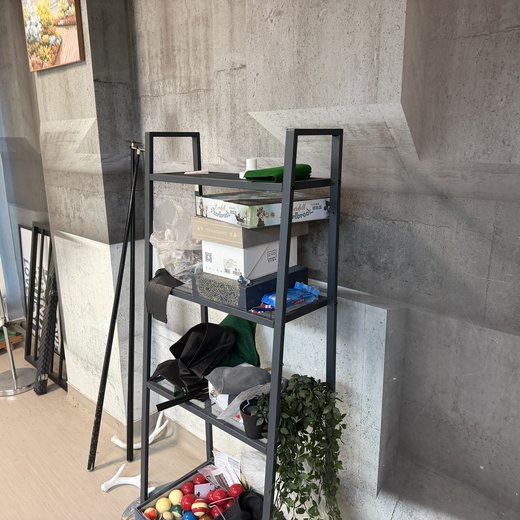}\\{\scriptsize Pos 4}\end{minipage}\hspace{1.5pt}%
  \begin{minipage}{0.155\linewidth}\centering\includegraphics[width=\linewidth]{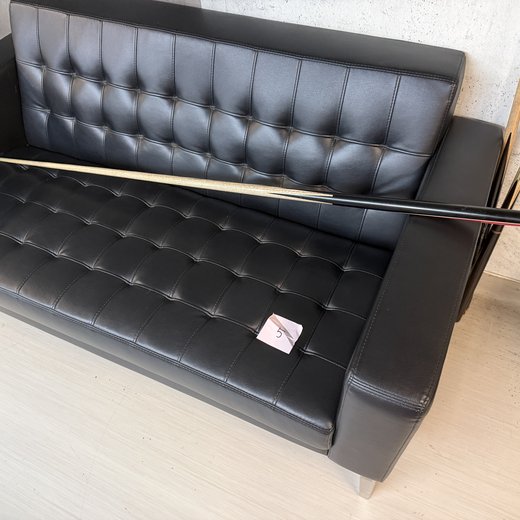}\\{\scriptsize Pos 5}\end{minipage}\hspace{1.5pt}%
  \begin{minipage}{0.155\linewidth}\centering\includegraphics[width=\linewidth]{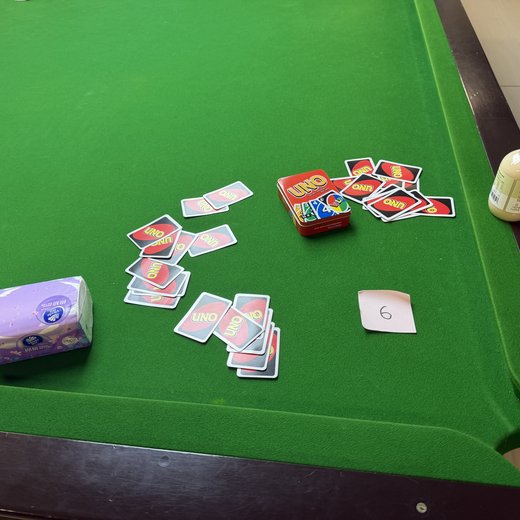}\\{\scriptsize Pos 6}\end{minipage}
  \caption{The two study rooms and their target positions. (a) L1, a conference room. (b) L2, an activity room. The lower pictures show the six predefined target positions in each room, each marked with a pink note. Both rooms were unfamiliar to all participants.}
  \label{fig:rooms}
  \Description{Two color photographs of the study rooms above two strips of small photographs. Photograph (a) shows L1, a conference room with purple curtains, a light wood floor, a white metal storage rack, small tables with folding chairs, and a whiteboard. Photograph (b) shows L2, a more open activity room with gray concrete walls, a green billiard table near the center, black leather sofas, and a corner shelf. The strip labeled L1 shows five close-up photographs of the six target positions in the conference room (positions four and five in one photo), each a surface such as a table, the floor, or a shelf with a small object and a pink marker note. The strip labeled L2 shows six close-up photographs of the target positions in the activity room, including a sofa, a chair, the billiard table, and a corner shelf, each with a small object and a pink marker note.}
\end{figure}

The study took place in two rooms unfamiliar to all participants (Figure~\ref{fig:rooms}). L1 was a conference room measuring approximately 39\,ft $\times$ 41\,ft (about 1{,}600\,ft$^2$); L2 was an activity room measuring approximately 31\,ft $\times$ 51\,ft (about 1{,}580\,ft$^2$). In C1 and C2 conditions, participants scanned each room themselves with the chest-mounted phone as they moved, and spoke to the system to locate the thermos cup.

\subsection{Study Design}
The study used a 3 (condition) $\times$ 2 (room) within-subjects design with two trials per cell: 12 trials per participant, 144 in total (Table~\ref{tab:design}). Trials were divided by rooms, six in one room and then six in the other, and within each room block the three conditions appeared in pairs of consecutive trials. Condition order, room order, and target positions followed a pre-generated counterbalanced schedule: six participants started in L1 and six in L2; within each room-order group, each of the six possible condition orders appeared once; and each participant was scheduled to encounter each of a room's six positions exactly once. Counterbalancing and per-trial target relocation were intended to mitigate order and learning effects as participants grew familiar with a room over its block.

\subsection{Procedure}
\label{sec:procedure}
Each session lasted roughly 1.5 to 2 hours. Two participants ran in parallel, one per room, so each visit accommodated two participants across the two rooms and two phones. The study was run by a fixed team following the same protocol in both rooms: an experimenter who placed the target and administered the conditions and a dedicated safety monitor.

\textbf{Clock-direction familiarization.} Before the trials, we asked each participant how familiar they were with clock-position (``o'clock'') referencing, then ran a short 2 minute tutorial we built to train the body-anchored clock convention that Touvigation uses. The tutorial had two stages: (1) orientation, in which the participant turns their body and a beep sounds when they face the correct clock direction; and (2) locomotion, in which, after orienting, they walk forward. Participants practiced both stages before the timed trials began, so that they were familiar with the C1 convention beforehand.

\begin{table}[t]
  \caption{Study design. Each participant completed 2 trials in every room--condition cell: 3 conditions $\times$ 2 rooms $\times$ 2 trials $=$ 12 trials per participant, and 144 trials across the 12 participants. Condition order and room order were counterbalanced (see text).}
  \label{tab:design}
  \small
  \begin{tabular}{@{}lll@{}}
    \toprule
    Room & Condition & Trials per participant \\
    \midrule
    L1 (conference room) & C1 Touvigation & 2 \\
                         & C2 Doubao      & 2 \\
                         & C3 Unassisted  & 2 \\
    \addlinespace
    L2 (activity room)   & C1 Touvigation & 2 \\
                         & C2 Doubao      & 2 \\
                         & C3 Unassisted  & 2 \\
    \bottomrule
  \end{tabular}
\end{table}

\textbf{Trials.} Before each trial, the experimenter placed the target at its scheduled position. In C1, the participant asked Touvigation by voice for the object to find, the thermos cup, and the system located it and guided the participant toward it. A nominal four-minute hard stop is applied. A tripod-mounted camera recorded each room from a third-person view and a dedicated safety monitor intervened only when necessary.

\subsection{Measures}
\label{sec:measures}
For RQ1, we measured, per trial, completion time and success. Completion time is the interval from trial start to the first correct touch of the target, timed to the nearest second from the synchronized session recordings and position logs. Success, and any touch of a wrong object, was adjudicated from the first-person audio and the third-person video. The neck-mounted device logged 20\,Hz position traces in all conditions, with per-trial metadata including the ground-truth target position in the room's map frame. First-person audio (system speech and participant speech) was recorded for each trial, and all sessions were video-recorded by the third-person cameras.

After each condition, participants completed a modified NASA-TLX~\cite{hart1988development} (for RQ2) and a customized questionnaires covering trust, perceived safety, and spatial awareness (for RQ3), and a short semi-structured interview (for qualitative study).  All items were administered and answered orally and captured in recordings.

\section{Results}
\label{sec:results}
This section reports findings for the three research questions. For RQ1, Touvigation produced the strongest overall task performance, with higher success and shorter failure-adjusted completion times than both baseline conditions. For RQ2, participants reported lower workload with Touvigation. For RQ3, Touvigation received higher ratings for spatial guidance, trust, perceived safety, and overall experience. 

We analyzed 144 trials from 12 participants (48 trials per condition) and all post-condition questionnaires. Friedman tests compared the three conditions; we report $\chi^2(2)$, $p$, and Kendall's $W$. Significant omnibus tests were followed by pairwise Wilcoxon signed-rank tests using normal approximations, with Holm correction across the three condition pairs.

\subsection{Timed Performance}

\textbf{Success.} All 48 Touvigation trials were successful, compared with 28 of 48 Doubao trials (58\%) and 41 of 48 unassisted trials (85\%), see Table ~\ref{tab:per-participant}. Participant-level success proportions differed across conditions, $\chi^2(2)=19.63$, $p<.001$, $W=.82$. Success proportions were higher with Touvigation than with Doubao ($p=.009$, $r=.90$) and unassisted search ($p=.020$, $r=.95$), and higher with unassisted search than with Doubao ($p=.009$, $r=.92$).

\begin{table}[t]
  \caption{Per-participant successes and mean completion time (s). Time$^{*}$ scores failures at 240\,s; Time$^{\dagger}$ counts successful trials only.}

  \label{tab:per-participant}
  \footnotesize
  \begin{tabular}{@{}l rr rrr rrr@{}}
    \toprule
    & \multicolumn{2}{c}{C1 Touvigation} & \multicolumn{3}{c}{C2 Doubao} & \multicolumn{3}{c}{C3 Unassisted} \\
    \cmidrule(lr){2-3}\cmidrule(lr){4-6}\cmidrule(lr){7-9}
    ID & Success & Time & Success & Time$^{*}$ & Time$^{\dagger}$ & Success & Time$^{*}$ & Time$^{\dagger}$ \\
    \midrule
    P1 & 4/4 & 101 & 1/4 & 206 & 106 & 2/4 & 140 & 41 \\
    P2 & 4/4 & 105 & 2/4 & 177 & 115 & 3/4 & 138 & 104 \\
    P3 & 4/4 & 62 & 1/4 & 230 & 200 & 3/4 & 133 & 98 \\
    P4 & 4/4 & 115 & 2/4 & 188 & 137 & 3/4 & 125 & 87 \\
    P5 & 4/4 & 75 & 4/4 & 119 & 119 & 4/4 & 94 & 94 \\
    P6 & 4/4 & 48 & 3/4 & 194 & 179 & 4/4 & 84 & 84 \\
    P7 & 4/4 & 60 & 2/4 & 151 & 62 & 3/4 & 99 & 52 \\
    P8 & 4/4 & 75 & 3/4 & 118 & 77 & 4/4 & 69 & 69 \\
    P9 & 4/4 & 68 & 3/4 & 138 & 104 & 4/4 & 107 & 107 \\
    P10 & 4/4 & 75 & 3/4 & 160 & 133 & 3/4 & 138 & 104 \\
    P11 & 4/4 & 71 & 2/4 & 195 & 151 & 4/4 & 78 & 78 \\
    P12 & 4/4 & 76 & 2/4 & 165 & 91 & 4/4 & 66 & 66 \\
    \midrule
    All & 48/48 & 78 & 28/48 & 170 & 120 & 41/48 & 106 & 83 \\
    \bottomrule
  \end{tabular}
\end{table}

\textbf{Completion time.}  Success differed across conditions. We first analyzed completion time with each failed trial scored at the 240\,s task limit (Time$^{*}$). Mean times were 77.5\,s (SD = 19.6) for Touvigation, 170.1\,s (SD = 35.0) for Doubao, and 105.9\,s (SD = 28.2) for unassisted search, with a significant difference across conditions, $\chi^2(2)=20.67$, $p<.001$, $W=.86$. Touvigation was faster than Doubao ($p=.007$, $r=.88$) and unassisted search ($p=.007$, $r=.79$), while unassisted search was faster than Doubao ($p=.007$, $r=.88$). When considering successful trials only (Time$^{\dagger}$), mean times were 77.5\,s (SD = 19.6), 122.6\,s (SD = 40.2), and 81.9\,s (SD = 21.3), respectively, $\chi^2(2)=15.17$, $p<.001$, $W=.63$. Touvigation and unassisted search were each faster than Doubao (both $p=.007$), whereas we found no evidence of a difference between Touvigation and unassisted search ($p=.530$, $r=.18$).

\begin{figure}[t]
  \centering
  \includegraphics[width=0.75\linewidth]{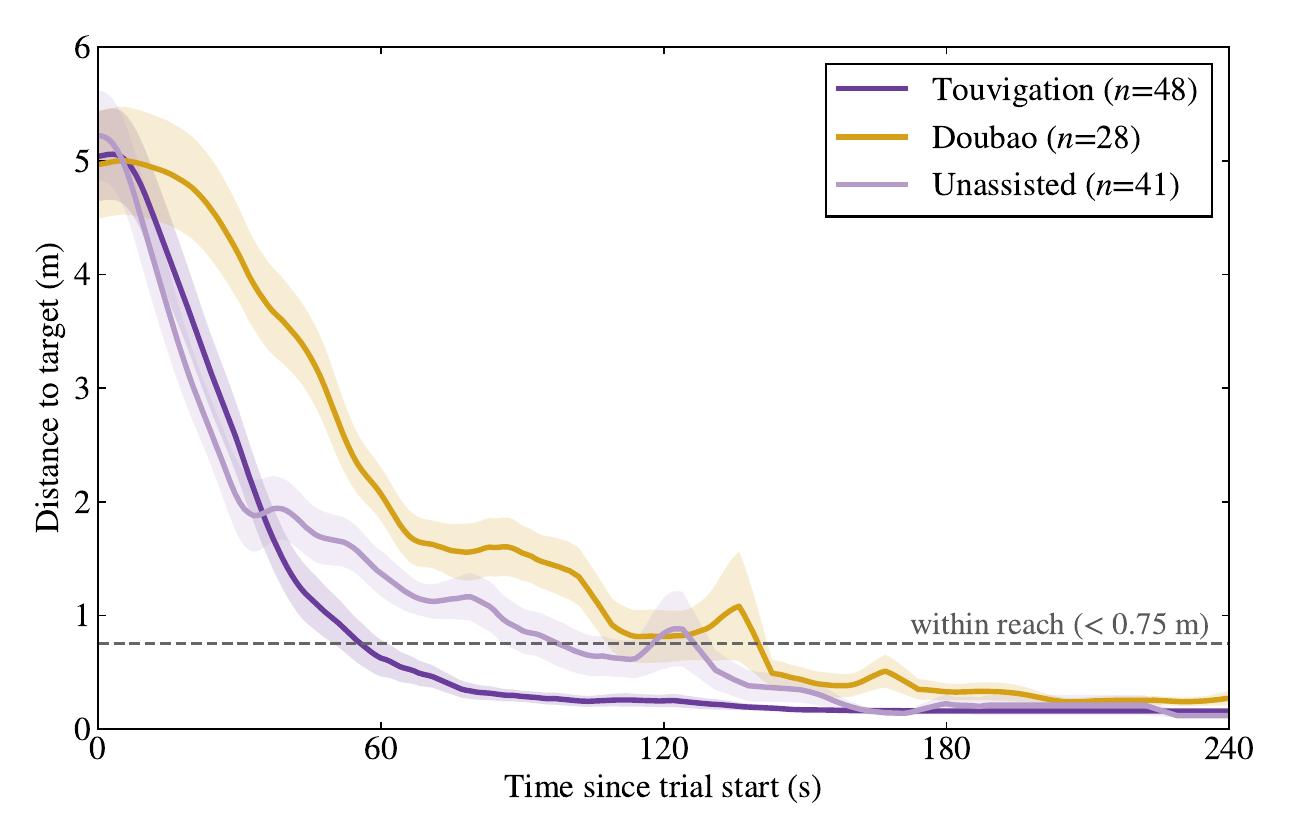}
 \caption{Mean straight-line distance to the target over time for successful trials with trajectory data, by condition. Shading shows $\pm$1 SE; within-reach zone ($<0.75$\,m)~\cite{qian2019portal,drillis1966body}.}
  \label{fig:distance-success}
  \Description{Line chart of mean distance to target against time for the three conditions, successful trials only. Touvigation falls into the within-reach zone by about one minute and unassisted search by about 100 seconds, overlapping thereafter; Doubao descends more slowly and enters the zone at about 140 seconds.}
\end{figure}

\textbf{How Participants Approached the Target} The 20 Doubao failures comprised 12 timeouts and 8 wrong-target attempts, whereas the 7 unassisted-search failures comprised 3 timeouts and 4 wrong-target outcomes. Figure~\ref{fig:distance-success} shows how participants approached the target during successful trials. The figure reports the mean straight-line distance remaining to the target at each second, with lower values indicating greater progress and the region below 0.75\,m indicating that the target was within reach~\cite{qian2019portal,drillis1966body}. Across successful trials, particpants were moving approximately 0.14\,m/s with Touvigation, 0.12\,m/s with Doubao, and 0.29\,m/s during unassisted search. We observed that participants moved fastest without assistance, as they did not pause for instructions. However, their rapid movement did not produce a consistently direct approach. The unassisted curve decreased rapidly at the first 30 seconds but then flattened near the target. Participants were slowing down and conducting a local tactile search once they reached the general area. Touvigation produced a similarly rapid reduction in target distance despite the lower speed, followed by a comparatively steady approach into the within-reach region. Doubao showed the slowest initial progress and even a temporary increase in distance at approximately 130\,s. While experimenting with Doubao, we observed that some participants moved away from the target after initially approaching it, because they were reorienting or searching around the target area when Duobao did not provide precise reaching instructions. In comparison, unassisted search supported faster movement, whereas Touvigation supported more direct and intuitive target acquisition toward the target.

\subsection{Workload}

\begin{figure}[t]
  \centering
  \includegraphics[width=\linewidth]{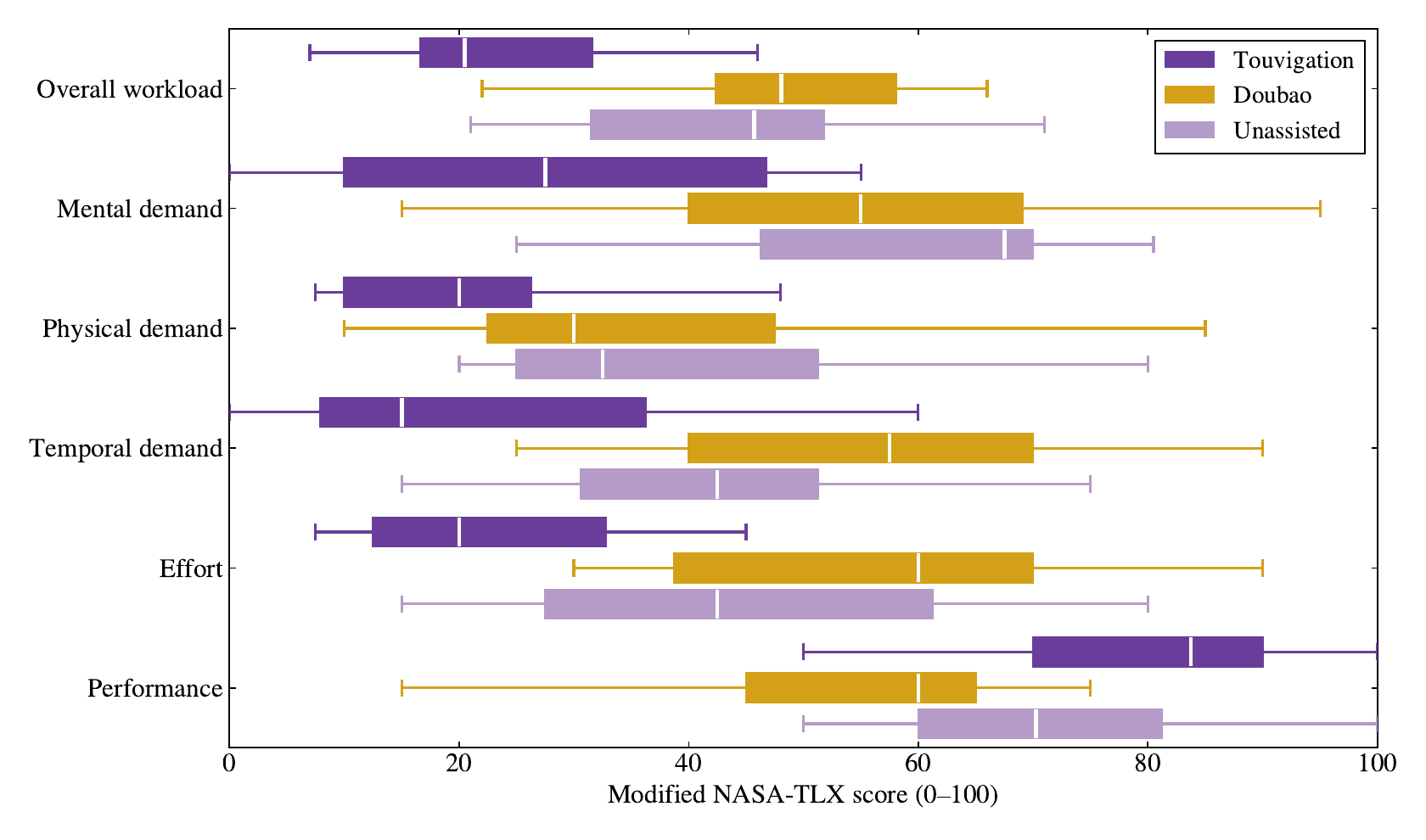}
  \caption{Participant-level distributions of overall workload and individual ratings from the NASA-TLX questionnaire across Touvigation, Doubao, and unassisted search (N = 12; 0–100 scale).}
  \label{fig:ratings}
  \Description{Horizontal boxplots compare the three conditions on overall workload, mental, physical, and temporal demand, effort, and self-rated performance, on a 0 to 100 scale. Touvigation, in purple, has the lowest boxes on every workload row and the highest box for performance. Unassisted search is highest on mental and physical demand while Doubao is highest on temporal demand and effort.}
\end{figure}

We assessed workload using the NASA-TLX and compared participant-level scores across conditions using Friedman tests followed by Holm-corrected pairwise Wilcoxon signed-rank tests. Overall workload means (SDs) were 24.2 (12.6) with Touvigation, 50.1 (16.4) with Doubao, and 43.5 (14.9) with unassisted search (Figure~\ref{fig:ratings}), $\chi^2(2)=10.50$, $p=.005$, $W=.44$. Workload was lower with Touvigation than with Doubao ($p=.009$, $r=.82$) and unassisted search ($p=.009$, $r=.86$). Mental and physical demand were lower with Touvigation than with both baselines (mental: $\chi^2(2)=17.15$, $p<.001$, $W=.71$; physical: $\chi^2(2)=11.02$, $p=.004$, $W=.50$). The Holm-corrected comparisons were $p=.013$, $r=.78$ and $p=.010$, $r=.89$ for mental demand, and $p=.015$, $r=.81$ and $p=.015$, $r=.89$ for physical demand, against Doubao and unassisted search, respectively. Touvigation--Doubao comparison remained significant within each subscale (temporal: $p=.020$, $r=.78$; effort: $p=.010$, $r=.89$; performance: $p=.022$, $r=.77$). 

\begin{figure}[t]
  \centering
  \includegraphics[width=\linewidth]{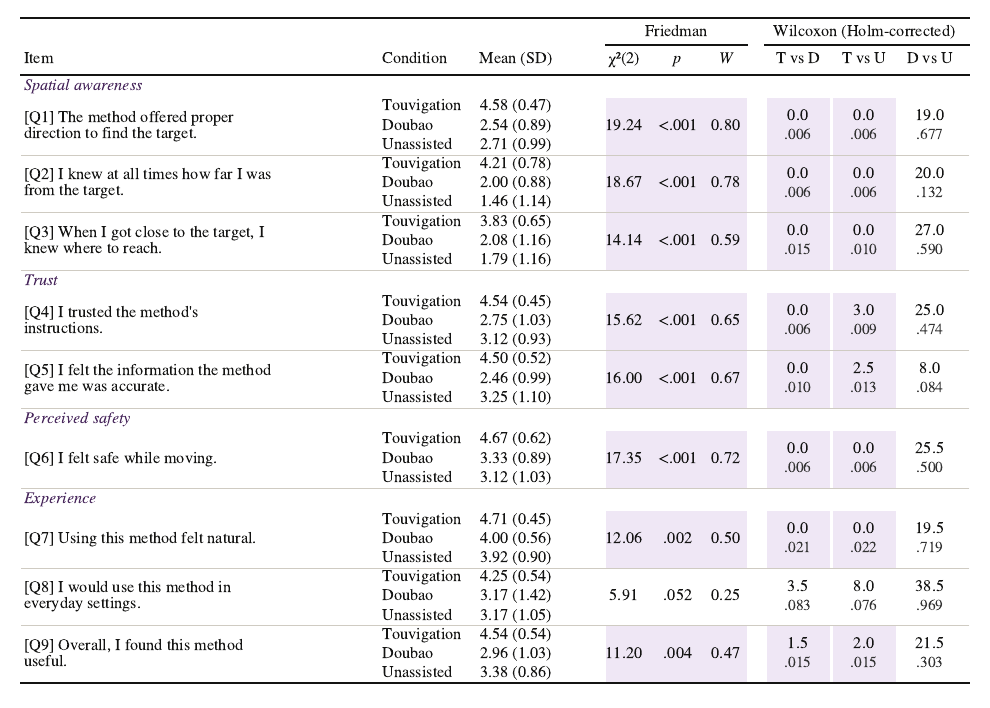}
  \caption{Experience items by condition: mean (SD) agreement on a 0--5 scale over the 12 participants, the Friedman test across the three conditions (Kendall's $W$ as effect size), and Wilcoxon signed-rank pairwise comparisons with Holm-corrected $p$ (T = Touvigation, D = Doubao, U = unassisted). Shaded cells are significant at $p < .05$.}
  \label{fig:experience}
  \Description{A table of nine questionnaire items grouped into spatial awareness, trust, perceived safety, and experience. For every item, Touvigation has the highest mean and its comparisons against Doubao and against unassisted search are significant, except for willingness to use the method in everyday settings. No comparison between Doubao and unassisted search is significant.}
\end{figure}

\subsection{Perceived Safety, Trust, and Spatial Awareness}

Figure~\ref{fig:experience} summarizes the nine experience items. Touvigation received higher ratings than both baselines for direction, distance, final reaching, trust, accuracy, perceived safety, naturalness, and usefulness (all Holm-corrected $p\leq.022$; Kendall's $W=.47$--$.80$). Willingness to use the method in everyday settings did not reach the significance threshold ($p=.052$). We found no evidence of a difference between Doubao and unassisted search on any item (all Holm-corrected $p\geq.084$).

\subsection{Qualitative Analysis}
We have coded the open questions in the questionnaires and interview script using both open and axial coding. Three coders separately coded the data and cross-verified the results. 

\subsubsection{Descriptions Supported Orientation but Required Translation into Action}

Despite Doubao's weaker task performance, participants continued to value it as a source of otherwise inaccessible visual information. Many participants described using the app whenever something needed to be seen and emphasized how
frequently it was used in everyday life (P7, P10, P11). For them, scene description remained useful even when it did not directly produce a successful route.

During the study, participants had to translate scene descriptions into movement and reaching actions. This became particularly difficult as their position or camera view changed, creating frequent conflicts between actions and the provided descriptions. Participants often could not identify the source of these conflicts: whether they resulted from (a) \textbf{network delays} (P3, P7), (b) \textbf{AI hallucinations}, where orientation and distance could be independently inaccurate (P6, P11), (c) \textbf{misinterpretation of guidance}, where participants understood the system's instructions differently from what was intended, or (d) \textbf{movement execution errors}, where participants' physical movements deviated from the instructed direction or distance (P1, P9).

Overall, participants agreed that Doubao supported scene awareness but placed the burden on them to determine whether descriptions were current, how they related to their bodies, and what actions to take next. Therefore, Doubao carried a heavier workload. Participants reported elevated mental demand (M = 54.8, SD = 23.4, vs. 28.9 with Touvigation), which they attributed to constantly verifying its answers against their own touch and movement. Temporal demand (M = 56.4, SD = 21.5) and effort (M = 56.9, SD = 19.7) were the highest of the three conditions for this reasons.

\subsubsection{Participants Built Search Plans Based on Individual Experience}

When guidance provided only coarse information, participants built their own search plans by combining these cues with prior knowledge of rooms, furniture, and object placement. For example, when told that a target was on a shelf, some inferred that the shelf was likely against a wall, located the wall first, and then searched along it (P3, P8, P9). Similarly, height cues such as ``at your abdomen'' helped participants infer likely surfaces, such as a table (P1, P6, P12). We observed participants using walls or large furniture to establish their position, tracing boundaries, and then narrowing the search area by hand (P1, P3, P5, P8, P9, P12).

This active search planning was reflected in participants' movement patterns. In the unassisted condition, participants moved fastest (0.29~m/s) and covered most of the distance within the first 30 seconds. However, unassisted search also produced the highest mental demand ($M=59.6$, $SD=18.7$) and physical demand ($M=40.4$, $SD=19.8$), consistent with participants having to construct and continuously update their own search plans while moving. We observed the effectiveness of these strategies varied across participants. Those who were congenitally blind or had been blind for years can often turned cues into effective search plans. However, when initial inferences failed, participants reverted to slower, systematic strategies, such as tracing surfaces inch by inch (P1, P5, P12). Rapid movement did not necessarily translate into faster target acquisition, as time gained during initial navigation could be lost during fine-grained search.

\begin{figure*}[t]
  \centering
  \includegraphics[width=\textwidth]{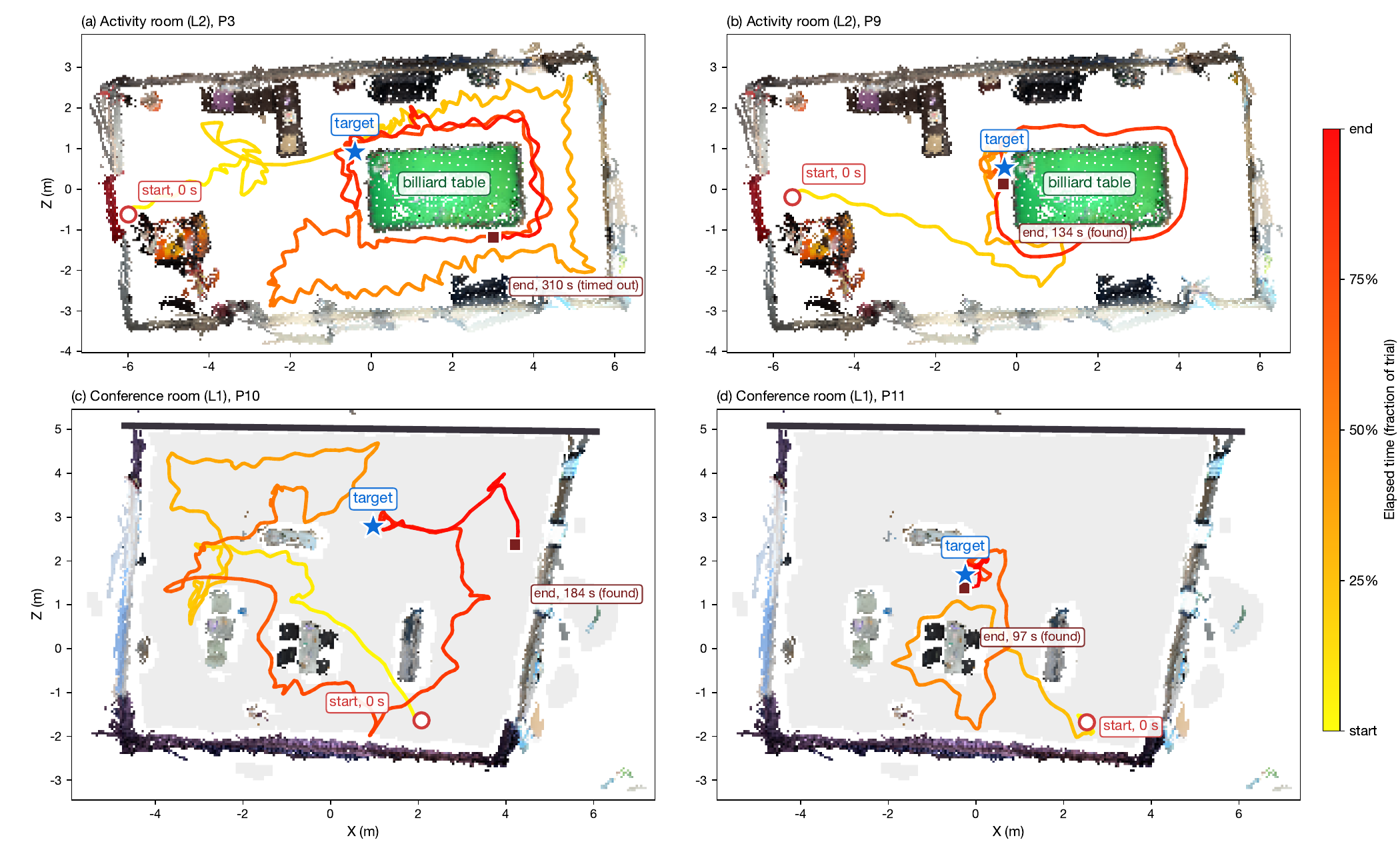}
  \caption{Four example unassisted-search trajectories overlaid on reconstructed room maps. Blue stars mark targets, circles mark starts, and squares mark endpoints. Path color encodes elapsed time normalized within each trial, from yellow at the start to red at the end. (a) P3 searched the activity-room perimeter and around the billiard table without confirming the target. (b) P9 passed near the target before finding it at 134 s. (c) P10 searched across the conference room before finding the target at 184 s. (d) P11 followed a more compact route and found the target at 97 s.}
  \label{fig:traj-unassisted}
  \Description{Four top-down room maps, two of the activity room and two of the conference room, each with a search path colored from yellow to red, a start circle, an end square, and a blue star marking the target.}
\end{figure*}

Moreover, systematic coverage did not indicate when participants had actually reached the target. Figure~\ref{fig:traj-unassisted} illustrates how this strategy unfolded across four unassisted trials. Participants used walls and large pieces of furniture to establish their position, then narrowed the search through boundary following and local tactile sweeps. P3 covered much of the activity-room perimeter and repeatedly circled the billiard table, but timed out without finding the target despite passing close to it several times. P9 also passed through the target area before completing a broad loop around the table and eventually finding it. In the conference room, P10 traversed much of the room before locating the target, whereas P11 followed a more compact path around the central tables. Together, these trajectories explain why unassisted search was generally reliable but exhaustive. Systematic coverage often led participants to the target, but did not provide a direct route or indicate precisely where to reach.

\subsubsection{Body-Anchored Cues Narrowed Search from Walking to Reaching}

With Touvigation, participants described guidance as a sequence of locally executable actions. Clock direction and step counts guided locomotion, while updates from each new position supported adjustment. P8 viewed immediate correction prompts as evidence that the system tracked their actions. P12 reported that a decreasing total number of steps to the target was a positive cue that they were moving in the right direction. As participants approached the target, guidance shifted from body-scale navigation to fine-grained reaching. Height cues (e.g., waist, knee, or ankle) helped participants determine where to position their arm and infer likely target surfaces (P5, P7, P11), while palm-scale distances specified the direction and extent of the final reach (P4, P5, P7, P1). Nearby-object warnings further supported the approach by identifying surrounding furniture (P12) and increasing participants' confidence in moving forward (P4, P9).  Guidance progressed across action scales: clock direction and steps supported locomotion, while height, palm-scale distance, and nearby-object information guided the final reach. This progression was also reflected quantitatively. Touvigation supported steady target approach. Participants rated Touvigation highest for knowing the target's direction (4.58), distance (4.21), and where to reach (3.83) on a 0--5 scale.

\subsubsection{Tactile Descriptions Made the Target Verifiable by Hand}

Touvigation's tactile descriptions gave participants an expectation of what the target should feel like before they touched it. P3 explained, ``After I found it, the feel in my hand said `this is the cup,' so the two confirm each other.'' Also, participant report that material and shape made this expectation discriminative (P3, P6, P7, P11). For example, P7 noted that knowing the target was metal helped her distinguishing other similar shaped objects by touching the surface. These properties allowed them to compare what they touched against and confirm if they had found the correct target.

Target dimensions also shaped how participants' search strategy. Knowing the approximate height and size allowed them to constrain their hand movement to where the target was likely to make contact. For example, P9 explained that knowing the cup was about 8 inches tall allowed him to raise his hand to approximately that height and sweep horizontally. At this level, his hand could contact the cup while passing above shorter objects on the same surface, reducing unnecessary contact and the risk of knocking them over. Tactile descriptions therefore supported both identifying the target after contact and making the search before contact more targeted and controlled.
\section{Discussion}\label{sec:discussion}

Our findings suggest that object acquiring in unfamiliar environments requires more than recognizing a target or describing its surroundings. Across the three research questions, Touvigation enabled more reliable and intuitive object acquisition, reduced workload, and improved participants' perceptions of spatial guidance, safety, and trust. Our qualitative findings further show that effective guidance needed to remain grounded in participants' changing bodily positions, become increasingly precise as they approached the target, and support continuous verification of their actions. We discuss these findings through three related interaction challenges: (1) translating scene understanding into action, (2) bridging walking and reaching, and (3) building confidence through continuous verification.

\subsection{From Scene Understanding to Actionable Object Finding}

For RQ1, Touvigation enabled more reliable object finding and shorter completion times with trajectories showing more sustained progress toward the target. With Doubao, participants received useful descriptions but still had to translate them into ``mental routes'' as their position and orientation changed. Unassisted search required similar planning. Participants transformed coarse placement cues into wall following, furniture-based localization, and progressively tactile actions. These strategies could bring participants near the target, but did not consistently specify how to approach or where to reach.

This contrast reveals an \textit{actionability gap} between knowing about a scene and acting. Conversational visual-assistance systems have expanded access to visual content (e.g. Be My AI), while interactive object-search systems have begun connecting recognition with user-directed search. Yet descriptions alone do not preserve their spatial meaning as users move. In our study, participants often supplied this transformation themselves by selecting landmarks, reconstructing spatial relations, checking whether descriptions remained current, and deciding where and how far to move. Touvigation instead maintained the spatial relationship between participant and target. The advantage was therefore not merely more precise information, but information that remained actionable through movement.

\subsection{The Last Meter of an Embodied Guidance Problem}

In unassisted search, participants could navigate toward the target nearby area fast, but often relied on trial-and-error to locate the exact object. Touvigation changes this experience by offering direct guidance to final acquisition, achieved through the two-stage embodied guidance. Indoor navigation systems commonly focus on reaching a destination or waypoint, but object acquisition requires support beyond arrival. Reaching the correct furniture or region does not ensure that the target can be found by hand.

The last meter was a critical stage where being near the target did not guarantee successful acquisition. P3 passed close to the target several times before timing out, while P10 swept just above the cup on the correct sofa. Even after contact, identity could remain uncertain; as P6 noted, ``I can only know I want to get a thermos, but how do I know if this thermos is the one I am looking for? I may need to call somebody to verify.'' Touvigation therefore supported not only reaching the target, but also precise hand-level localization and tactile confirmation. As P4 described, ``I moved my left hand two palms to the left and touched it.''

\subsection{Confidence to Act Under Physical and Social Risk}

For RQ3, participants rated Touvigation higher in perceived safety, and trust. One possible explanation is that our system supported participants' ability to verify guidance through actions, and sometimes \textit{not} through actions. This is particularly important because object acquiring actions (especially in public) contains a social component. Participants described feeling socially exposed when searching in front of others, where moving uncertainly, repeatedly ``groping around'', or touching the wrong object could appear awkward (P4, P5, P7, P12). As P7 explained, ``groping around in front of colleagues felt rather awkward and unpleasant.'' Uncertainty therefore carried not only \textbf{physical} and \textbf{cognitive costs}, but also \textbf{social costs}.

Touvigation offered participants' ability to proceed despite this uncertainty, we view this as \textit{confidence to act}.  Participants perceive real-time feedback as they moved, receive corrections after deviations, and confirm the target through touch. P8 gained confidence when the system immediately detected and corrected a wrong turn, while P4 reported that she ``dared to walk boldly'' when guidance consistently matched the environment. Confidence thus build through repeated loops of guidance, action, and verification. In contrast, With Doubao, participants could not always determine whether a mismatch came from the system, their own movement, or their interpretation. 

Another benefit for continuous guidance is that users no longer carries the burden of detecting and recovering from errors (As reflected from the TLX scoring). For embodied experience, reliability should thus be considered not only in terms of how often a system is correct, but also in terms of how safely, privately, and confidently users can recognize and recover when it is wrong.


\section{Conclusion}

We present Touvigation, an embodied object-acquisition system that guides blind and low-vision users from locating a distant target to reaching and confirming it by hand. Touvigation maintains a persistent semantic-spatial representation of the target and continuously re-anchors guidance to the body: clock directions support orientation, personalized step counts support locomotion, and body-relative height, palm-scale distance, and tactile properties support final acquisition. Our evaluation with blind participants showed that closing this action-feedback loop made acquisition more intuitive and direct, reduced workload, and improved spatial awareness, trust, and perceived safety. These findings shift the design goal of AI visual assistance from producing increasingly detailed descriptions to maintaining guidance that remains actionable as users move. By allowing users to verify progress, recover from deviations, and confirm targets through touch, Touvigation demonstrates how embodied assistive AI can foster confidence to act while reducing the physical risks and social discomfort associated with uncertain searching.
\section{Limitations and Future Work}

Our participants were blind adults classified at Level~1 and were recruited through one regional community. Their experiences may not represent the wider diversity of blind and low-vision people, including people with residual vision, different onset histories, and different orientation and mobility practices. The study also captured short-term use following familiarization. Longitudinal field deployments are needed to examine how users learn the guidance, calibrate confidence in the system, integrate it with established mobility aids, and experience it in social settings where uncertain searching may attract unwanted attention.

Touvigation currently depends on a LiDAR-equipped, chest-mounted iPhone, an initial room scan, stable spatial tracking, and cloud-based semantic processing. Layout changes, moving objects, occlusion, tracking drift, poor lighting, or network delays could make the target or semantic information inaccurate. Future work should support dynamic map updates, communicate uncertainty, provide recoverable guidance when tracking fails, reduce dependence on cloud processing, and evaluate privacy-preserving deployment outside controlled environments.

\section{Acknowledgement}

We used AI tools to assist with system coding and grammar checking in this paper. We thank everyone who helped coordinate the user studies, and especially all BLV participants for their time, participation, and valuable feedback.

\bibliographystyle{ACM-Reference-Format}
\bibliography{references}

\end{document}